\documentclass[twocolumn,floats,floatfix,showpacs,prd,superscriptaddress,nofootinbib]{revtex4-1}

\usepackage{graphicx,epsfig}
\usepackage{amssymb,amsmath,amsthm,amsfonts}
\usepackage{bm}
\usepackage[inline]{enumitem}
\usepackage{tensor}
\usepackage[linktocpage]{hyperref}
\usepackage[caption=false]{subfig}
\usepackage[usenames,dvipsnames]{xcolor}
\usepackage{url}
\usepackage[inline]{enumitem}
\usepackage{xspace}
\usepackage{comment}
\usepackage[normalem]{ulem}

\newcommand{\tder}{\partial_t}
\newcommand{\rder}{\partial_r}
\newcommand{\Rder}{\partial_R}
\newcommand{\OO}[1]{\mathcal{O}\left(#1\right)}

\definecolor{brightlavender}{rgb}{0.75, 0.58, 0.89}

\newcommand{\sapienza}{Dipartimento di Fisica, Sapienza Università 
	di Roma, Piazzale Aldo Moro 5, 00185, Roma, Italy}
\newcommand{\infn}{INFN, Sezione di Roma, Piazzale Aldo Moro 2, 00185, Roma, Italy}

\newcommand{\CC}{C\nolinebreak\hspace{-.05em}\raisebox{.4ex}{\tiny\bf +}\nolinebreak\hspace{-.10em}\raisebox{.4ex}{\tiny\bf +}}

\begin{document}

\title{Black hole spectroscopy and nonlinear echoes in Einstein-Maxwell-scalar theory}
\author{Marco Melis}
\email{marco.melis@uniroma1.it}
\affiliation{\sapienza}
\affiliation{\infn}

\author{Fabrizio Corelli}
\email{fabrizio.corelli@uniroma1.it}
\affiliation{\sapienza}
\affiliation{\infn}

\author{Robin Croft}
\email{robin.croft@uniroma1.it}
\affiliation{\sapienza}
\affiliation{\infn}

\author{Paolo Pani}
\email{paolo.pani@uniroma1.it}
\affiliation{\sapienza}
\affiliation{\infn}

\begin{abstract}
    In the context of Einstein-Maxwell-scalar theory with a nonminimal coupling between the electromagnetic and scalar field, we study linear (non)radial perturbations and nonlinear radial dynamics of spherically symmetric black holes.
    In a certain region of the parameter space, this theory admits hairy black holes with a stable photon sphere. This has a counterpart in the effective potential of linear perturbations, featuring multiple maxima and minima.
    The corresponding quasinormal mode spectrum contains long-lived modes trapped in the potential cavity and the time-domain linear response displays echoes, as previously observed for horizonless compact objects.
    Interestingly, the black-hole dynamics in this theory can be studied at the nonlinear level. By performing fully-fledged 1+1 simulations, we show that echoes are present even when the nonlinearities are significant.
    To our knowledge, this is the first example of echoes appearing in a consistent theory beyond a linearized analysis.
    In a follow-up work we will study whether this feature is also present in the post-merger signal from black hole collisions in this theory.
\end{abstract}

\maketitle

\tableofcontents
\section{Introduction}
The gravitational-wave~(GW) signal emitted after the merger of two black holes~(BHs) provides invaluable information about the underlying gravitational dynamics and the nature of the remnant as the latter relaxes toward a stationary configuration.
In the post-merger (ringdown) phase, this signal is modelled as a superposition of the remnant quasinormal modes~(QNMs)~\cite{Berti:2009kk,ringdownreview}, damped sinusoids associated with the excitation of the photon sphere of the BH remnant~\cite{Ferrari:1984ozr,Cardoso:2008bp}.
This intuitive correspondence holds in General Relativity~(GR) due to the particularly simple effective potential for BH perturbations, featuring a single maximum near the photon sphere and decaying to zero near the horizon and at infinity.

Small continuous deformations of this potential introduce slight corrections to the QNMs. This is what happens in extensions of GR, especially in effective field theories modifying the Einstein-Hilbert action by perturbatively small terms.

On the other hand, more drastic changes introduced by matter overdensities at large distance~\cite{Barausse:2014pra,Cheung:2021bol} or at the horizon scale (as predicted in certain quantum-gravity scenarios~\cite{Cardoso:2017cqb,Bena:2022rna} and by dark matter~\cite{Bertone:2024wbn} or accretion disk~\cite{Cardoso:2020nst,Cannizzaro:2024yee} models) modify the ringdown picture significantly.
Indeed, if the effective potential develops a cavity, the QNM spectrum is drastically modified. This can occur either because of reflective inner boundary conditions (as in the case of ultracompact horizonless objects~\cite{Cardoso:2016rao,Cardoso:2017cqb}), or because of matter bumps~\cite{Barausse:2014pra,Cheung:2021bol,Rosato:2024arw,Ianniccari:2024ysv} or even in the pure-vacuum BH case if the effective potential features several peaks due to beyond-GR effects.
In all these cases the QNM spectrum would contain long-lived modes associated with low-frequency perturbations trapped within the cavity, which can escape due to wave tunneling~\cite{Cardoso:2019rvt}.
Furthermore, the effective potential develops a minimum, which roughly corresponds to a \emph{stable} photon sphere, potentially prone to nonlinear instabilities~\cite{Cardoso:2014sna,Cunha:2017qtt}.

In recent years there has been considerable interest in understanding and modelling the ringdown of ultracompact objects displaying the above features, with the motivation of developing novel tests of gravity in the post-merger phase.
While the time-domain signal is dominated by the low-frequency cavity modes at late times, the prompt ringdown is mostly dominated by the shape of the potential near the unstable photon sphere, and is unaffected by the cavity if the latter is sufficiently wide~\cite{Cardoso:2019rvt}.
At intermediate times, the signal features repeated ``echoes''~\cite{Cardoso:2016rao,Cardoso:2016oxy,Cardoso:2017cqb} associated with the trapped modes reflected back and forth in the cavity and slowly tunneling out.
Unfortunately, this interesting phenomenon has been mostly studied for linearized perturbations of ad-hoc simplified models~\cite{Cardoso:2016rao,Cardoso:2016oxy,Wang:2019rcf,Ferrari:2000sr,Raposo:2018rjn,Pani:2018flj,Mannarelli:2018pjb,Zhang:2017jze,Oshita:2018fqu}.
Even when considering a more theoretical-motivated top-down approach (as in the case of fuzzballs~\cite{Mathur:2005zp,Bena:2007kg,Balasubramanian:2008da,Bena:2022rna}) the dynamics have been studied only at the linearized level~\cite{Ikeda:2021uvc,Heidmann:2023ojf,Dima:2024cok}.
This is due to the formidable challenge of finding consistent models of ultracompact objects that would be amenable for nonlinear evolution (see~\cite{Dailey:2023mvn,Dailey:2024kjg,Ma:2022xmp,Deppe:2024qrk} for some recent effective frameworks to perform nonlinear simulations of ultracompact objects, and~\cite{Siemonsen:2024snb} for merger simulations of spinning boson stars with stable light rings).

In this context, the main motivation for this paper is twofold.
First, we wish to study the ringdown of BHs in a theory that can yield large deviations from GR, possibly producing multi-peaked effective potentials for vacuum BH spacetimes. This automatically excludes effective field theory approaches where GR deviations are parametrically small. Second, we wish to focus on a theory with a well-posed initial value formulation for any value of the coupling constants, since the ringdown notably describes the final stage of a binary coalescence that should be simulated by evolving the full set of nonlinear evolution equations.
These two requirements (large deviations from GR and well-posedness) exclude many modified gravity theories and go beyond many previous studies. While there has been significant progress in the last few years regarding the well-posed formulation of theories beyond GR (see, e.g.,~\cite{Figueras:2024bba} for a recent result and~\cite{Foucart:2022iwu} for a review), this is typically limited to effective field theories that by construction ought to provide perturbatively small corrections to GR. Even in those cases in which simulations can be performed beyond the effective-field-theory regime, pathologies like elliptic regions appear~\cite{Ripley:2019irj,Corelli:2022pio,Corelli:2022phw,Hegade,Thaalba:2024htc}, casting doubts on the validity of such theories at large couplings.

Here we focus on Einstein-Maxwell-scalar~(EMS) theory, describing a real scalar field $\phi$ minimally coupled to Einstein's gravity and non-minimally coupled to electromagnetic Maxwell's field:
\begin{equation}\label{eq:EMS_action}
    S = \frac{1}{16\pi} \int d^4x \sqrt{-g}\,[R - 2\partial_\mu\phi \partial^\mu\phi - F[\phi]F_{\mu\nu}F^{\mu\nu}] \;,
\end{equation}
where units are set in such a way that $c = G = 4 \pi \epsilon_0 = 1$, $g_{\mu\nu}$ is the metric tensor, $R$ is the Ricci scalar and $F_{\mu\nu}$ is the usual Maxwell field strength tensor expressed in terms of the vector potential $A_\mu$, as $F_{\mu\nu} = \partial_\mu A_\nu - \partial_\nu A_\mu$. Additionally $F[\phi]$ is the generic coupling function between the scalar field and the Maxwell invariant $F_{\mu\nu}F^{\mu\nu}$.

This model is particularly motivated and appealing because: 
\begin{enumerate}
  \renewcommand{\labelenumi}{\roman{enumi}.}
  \item The gravitational sector is described by GR, and hence we expect strong hyperbolicity for any reasonable choices of $F[\phi]$;
  \item It admits hairy BH solutions~\cite{Herdeiro:2018wub,Fernandes:2019rez,Blazquez-Salcedo:2020nhs} that can be significantly different from the Reissner-Nordstrom~(RN) ones;
  \item Some of these hairy BHs feature a multipeaked potential~\cite{Guo:2022umh,Guo:2024cts}, at least for some perturbations\footnote{We note that, although \cite{Guo:2022umh} discusses multipeaked potentials for scalar perturbations of a hairy BH, the analysis in that work appears to be wrong, since it neglects the unavoidable coupling between scalar and gravitational perturbations in a hairy BH background~\cite{Myung:2019oua}, as also discussed below.}.
\end{enumerate}

Binary BH mergers in this theory with $F[\phi]\propto e^{-2 \alpha \phi}$, where $\alpha$ is a dimensionless coupling constant, have been studied in Ref.~\cite{Hirschmann:2017psw}, motivated by computing the full merger signal in a theory that can be nonperturbatively different from GR.
The QNMs of weakly-charged BHs in this theory were studied in \cite{Brito:2018hjh}.
Here we will consider a different coupling function, giving rise to BH scalarization~\cite{Herdeiro:2018wub,Fernandes:2019rez} (see~\cite{Doneva:2022ewd} for an overview) and more striking deviations from GR.

\section{EMS action and scalarized BH solutions}

Varying the action~\eqref{eq:EMS_action} with respect to $\phi$, $A_\mu$ and $g_{\mu\nu}$ we derive the following field equations
\begin{align}
\begin{split}
    \Box\phi - \frac14 \frac{\delta F[\phi]}{\delta \phi} F_{\mu\nu}F^{\mu\nu} &= 0 \;, \label{eq:scalar_eq}
    \end{split}\\
\begin{split}
    \nabla_{\mu}(F[\phi]F^{\mu\nu}) &= 0 \;,\label{eq:EM_eq}
    \end{split}\\ 
\begin{split}
    R_{\mu\nu} - \frac12 R \, g_{\mu\nu} &= T_{\mu\nu}^{\phi} + T_{\mu\nu}^{\text{EM}} \label{eq:Einstein_eq}\;, 
    \end{split}
\end{align}
where $T_{\mu\nu}^{\phi}$ and $T_{\mu\nu}^{\text{EM}}$ are the stress-energy tensors of the scalar and electromagnetic fields, respectively,
\begin{align}
    T_{\mu\nu}^{\phi} &= 2 \nabla_\mu\phi \nabla_\nu\phi - g_{\mu\nu}\nabla_\rho\phi \nabla^\rho\phi \;,\\
    T_{\mu\nu}^{\text{EM}} &= (2 F_{\mu\rho} F_{\nu}{}^\rho - \frac12 g_{\mu\nu} F_{\rho\sigma}F^{\rho\sigma})F[\phi] \;.
\end{align}

\subsection{Spherically symmetric BH solutions}
Following \cite{Herdeiro:2018wub}, static and spherically symmetric BH solutions to this theory can be found by imposing the ansatz
\begin{equation}\label{eq:bgBH_sol}
    ds^2 = -N(r) e^{-2\delta(r)} dt^2 + \frac{1}{N(r)}dr^2 + r^2(d\theta^2 + \sin^2\theta d\varphi^2) \;,
\end{equation}
where, using the Misner-Sharp mass function $m(r)$, $N(r)=1-2m(r)/r$~\cite{PhysRev.136.B571}. Spherical symmetry also implies that the electromagnetic field $A_\mu = (V(r),0,0,0)$ and the scalar field $\phi(r)$ only depend on the radial coordinate $r$. With these assumptions the field equations can be simplified as follows:
\begin{align}\label{eq:ode1}
    \delta' + r\phi'^2 &= 0 \;, \\ \label{eq:robin-d2Vdr2}
    (e^{\delta} F[\phi] \, r^2 V')' &= 0 \;, \\
    r(r-2m)\phi'^2 + r^2 V'^2 e^{2\delta} F[\phi] -2m' &= 0 \;, \\
    \nonumber r(r-2m)\phi'' - [2(m+rm'-r)+(r^2-&2mr)\delta']\phi' \\ 
    + \frac{r^2 V'^2 e^{2\delta}}{2}\frac{\delta F[\phi]}{\delta \phi}&= 0 \;,
\end{align}
where the prime denotes a derivative with respect to $r$. Eq.~\eqref{eq:robin-d2Vdr2} can be integrated with respect to $r$ giving
\begin{equation}\label{eq:ode4}
    V' = -\frac{Q}{r^2 F[\phi] \, e^\delta} \;,
\end{equation}
where the integration constant $Q$ is the electric charge of the spacetime.
In order to solve the ordinary differential equations~\eqref{eq:ode1}-\eqref{eq:ode4} we consider the following expansion near the BH horizon at $r=r_H$:
\begin{align}
    m(r) &= \frac{r_H}{2} + m_1(r-r_H) + \dots \;,\\
    \delta(r) &= \delta_0 + \delta_1 (r-r_H) + \dots \;, \\
    V(r) &= V_1 (r - r_H) + \dots \;,\\
    \phi(r) &= \phi_0 + \phi_1 (r -r_H) + \dots \;,
\end{align}
where
\begin{align}
    m_1 &= \frac{Q^2}{2 \, r_H^2 F[\phi_0]} \;,\\
    \delta_1 &= -\phi_1^2 r_H \;, \\
    \phi_1 &= \frac{Q^2}{2 \, r_H Q^2 F[\phi_0] - 2 r_H^3 F[\phi_0]^2} \frac{\delta F[\phi_0]}{\delta \phi} \;,\\
    V_1 &= - \frac{Q^2}{r_H^2 e^{\delta_0} F[\phi_0]} \;.
\end{align}
The two parameters $\delta_0$ and $\phi_0$ are found with a shooting method, imposing that the solution is asymptotically flat and given by the following expansion
\begin{align}
    m(r) &= M - \frac{Q^2 + Q_s^2}{2 r} + \dots \;, \\
    \delta(r) &= \frac{Q_s^2}{2 r^2} + \dots \;,\\
    \phi(r) &= \frac{Q_s}{r} + \frac{Q_s M}{r^2} + \dots \;, \\
    V(r) &= V_\infty + \frac{Q}{r} + \dots \;,
\end{align}
where $M$ is the ADM mass, $Q_s$ is the charge of the scalar field, and $V_\infty$ is the electrostatic potential at infinity.

\begin{figure}
  \centering
  \includegraphics[width=\columnwidth]{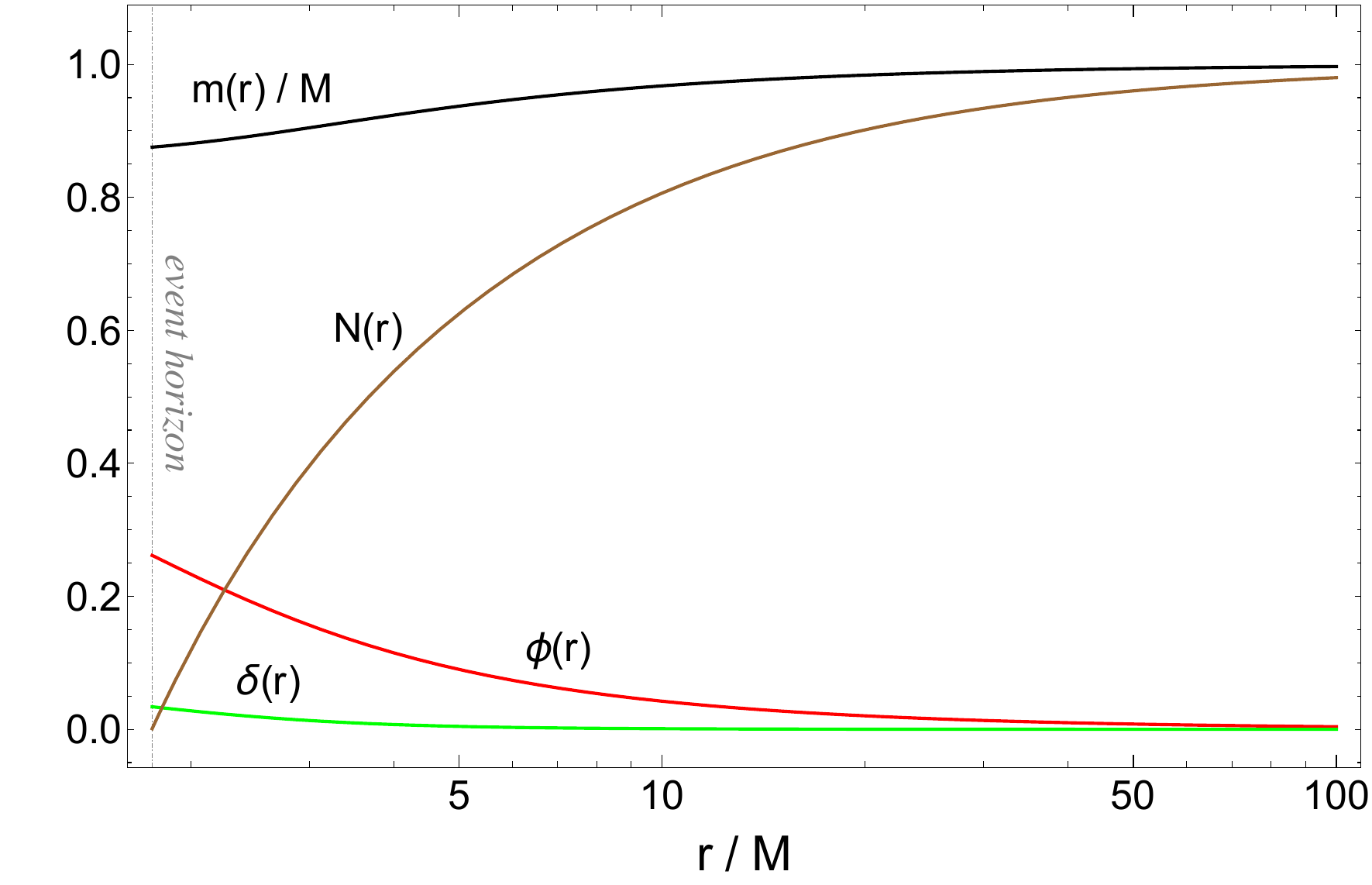}
  \caption{Metric functions $m(r)$, $N(r)$, $\delta(r)$ and scalar field $\phi(r)$ of the scalarized BH with $\alpha = 20$ and $q = Q/M = 0.7$.
  }
  \label{fig:scal_sol}
\end{figure}

For concreteness, we consider the coupling function 
\begin{equation}
    F[\phi] = e^{\alpha \phi^2}\,, \label{coupling}
\end{equation}
where $\alpha \geq 0$ is a dimensionless coupling constant.
The corresponding EMS model admits the scalar-free solution given by the RN BH with mass $M$ and electric charge $Q$, for any value of $\alpha$. This corresponds to $m(r) = M - \frac{Q^2}{2 r}$, $V(r)=V_\infty+Q/r$, and $\phi=\delta=0$. For any $\alpha > 1/4$ the theory also admits scalarized BH solutions with a charge $Q$ ranging from a minimum value (at which they connect to the RN BHs) to a maximum critical value $Q_\text{crit}$ marking the upper bound of the domain of existence~\cite{Herdeiro:2018wub}. In the following we will focus only on the so-called fundamental modes of scalarized BHs, namely the configurations for which the scalar field $\phi(r)$ has zero nodes.
As an example of scalarized solution, in Fig.~\ref{fig:scal_sol} we display the metric functions and the scalar field for $\alpha = 20$ and $q = Q/M = 0.7$. 

\subsection{Photon spheres}
In order to find the photon sphere(s) of these background solutions we study the geodesics described by massless particles in the spacetime of the scalarized BH.  Solving the geodesic equation for a photon is equivalent to solving the Euler–Lagrange equations 
\begin{equation}\label{eq:ELeqs}
    \frac{\partial \mathcal{L}}{\partial x^\rho} - \frac{d}{d\lambda} \frac{\partial \mathcal{L}}{\partial \Dot{x}^\rho} = 0 \,,
\end{equation}
where the dot denotes derivation with respect to the affine parameter $\lambda$, and $\mathcal{L}$ is the Lagrangian of a massless particle in the metric, given by Eq.~\eqref{eq:bgBH_sol}
\begin{equation}
    \mathcal{L} = \frac{1}{2}\left[ -N e^{-2 \delta} \Dot{t}^2 + \frac{1}{N}\Dot{r}^2 + r^2 \Dot{\theta}^2 + r^2 \sin^2{\theta} \Dot{\varphi}^2 \right].
\end{equation}
Since the spacetime is spherically symmetric we can restrict ourselves to the study of planar orbits, setting $\theta = \pi / 2$ without loss of generality.
From Eq.~\eqref{eq:ELeqs} we derive the equations for the $t$, $\varphi$ and $r$ components of the null geodesic
\begin{equation}
\Dot{t} = \frac{E}{N e^{-2 \delta}} \,, \quad
\Dot{\varphi} = \frac{L}{r^2} \,, \quad
\Dot{r}^2 = e^\delta [E^2 - V_{\rm ph}(r) L^2] \,,
\end{equation}
where $E$ and $L$ are the constants of motion associated to the two isometries of the spacetime, and the potential $V_{\rm ph}$ is given by 
\begin{equation}\label{eq:pot_ph}
    V_{\rm ph}(r) = \frac{N e^{-2 \delta}}{r^2} \,.
\end{equation}
The presence of a photon sphere at radial distance $\bar{r}$ is defined by null radial velocity and acceleration ($\Dot{r}=\ddot{r}=0$), that translate into the equations $E^2 = V_{\rm ph}(\Bar{r}) L^2$ and $d V_{\rm ph}(\Bar{r}) / dr = 0$. Local maxima (minima) of $V_{\rm ph}$ correspond to unstable (stable) photon orbits. 

\begin{figure}[h]
  \centering
  \includegraphics[width=\columnwidth]{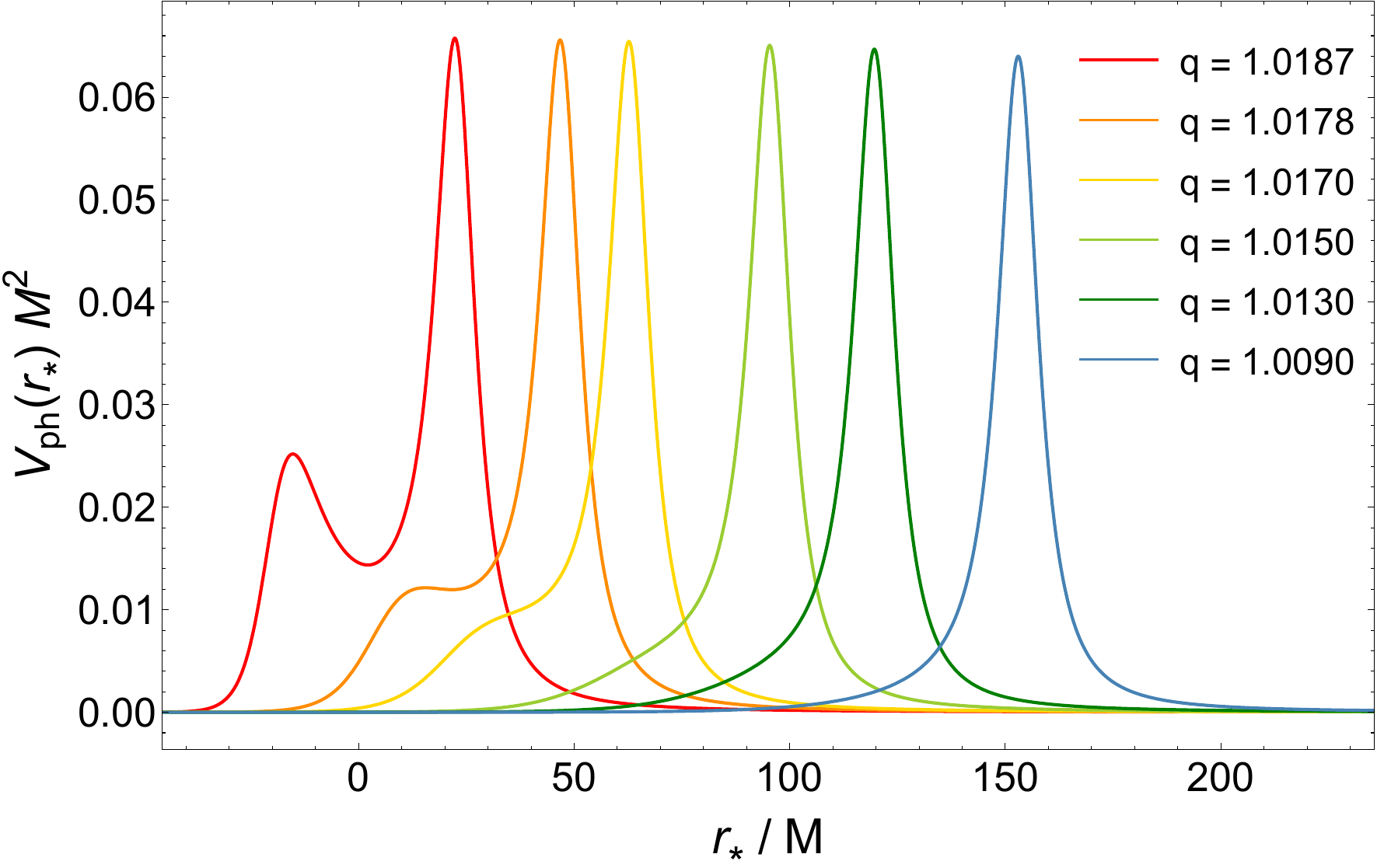}
  \caption{Null-geodesic potential $V_{\rm ph}$ in terms of the tortoise coordinate $r_*$, for different values of $q = Q/M$ in the case $\alpha = 0.6$. The potential for $q = 1.0187$ shows a well, signaling the presence of a stable photon sphere.} 
  \label{fig:Vph_a06}
\end{figure}

The parameter space of scalarized BH solutions reveals a region where a stable photon sphere appears. The latter is present for small values of the coupling constant $\alpha$ and near the critical charge $Q_\text{crit}$, above which the solutions cease to exist~\cite{Gan:2021xdl,Gan:2021pwu}.
A representative example is shown in Fig.~\ref{fig:Vph_a06}, where we display the profile of the potential in Eq.~\eqref{eq:pot_ph} as a function of the tortoise coordinate $r_*$, defined by $dr_*/dr = e^{\delta}/N$. We considered the case $\alpha = 0.6$, and different values of the charge-to-mass ratio $q$. As we can see $V_{\rm ph}$ develops a stable photon sphere as $q \to q_\text{crit} \approx 1.0187$.

\section{Perturbed field equations}
Motivated by the existence of stable photon spheres in certain hairy BHs in this theory, we move on and study linear perturbations of these solutions.
\subsection{Radial perturbations}
Let us start with the simple case of purely radial perturbations on the background solutions described by Eq.~\eqref{eq:bgBH_sol}, keeping the coupling function $F[\phi]$ general.  Following~\cite{Fernandes:2019rez} we consider spherically symmetric, linear perturbations of the metric, vector potential, and scalar field,
\begin{align}\label{eq:lin_spher_ansatz}
    ds^2 &= -\Tilde{N}(t,r) \, e^{-2 \Tilde{\delta}(t,r)} dt^2 \nonumber \\
    & \hspace{1.5cm}+ \frac{dr^2}{\Tilde{N}(t,r)} + r^2(d\theta^2 + \sin^2\theta d\varphi^2) \;, \\
    A &= \Tilde{V}(t,r) dt\;,\\
    \phi &= \Tilde{\phi}(t,r) \;,
\end{align}
where 
\begin{align}
    \Tilde{N}(t,r) &= N(r) + \epsilon \, N_1(t,r)\;, \\
    \Tilde{\delta}(t,r) &= \delta(r) + \epsilon \, \delta_1(t,r)\;, \\
    \Tilde{\phi}(t,r) &= \phi(r) + \epsilon \, \phi_1(t,r)\;, \\
    \Tilde{V}(t,r) &= V(r) + \epsilon \, V_1(t,r)\;.
\end{align}
Using the field equations, and keeping only terms linear in $\epsilon$, it is possible to derive the following relations between the perturbation functions:
\begin{align}
    \Dot{N}_1(t,r) &= -2 r N(r) \phi'(r) \Dot{\phi}_1(t,r) \;, \\
    \delta_1'(t,r) &= -2 r \phi'(r) \phi_1'(t,r) \;,\\
    \Dot{V}_1'(t,r) &= -V'(r)\biggl( \Dot{\delta}_1(t,r) + \frac{\Dot{\phi}_1(t,r)}{F[\phi]} \frac{\delta F[\phi]}{\delta \phi} \biggr) \;,
\end{align}
where the prime and the dot denote partial derivatives with respect to $r$ and $t$, respectively. Through $t$ integration, the above equations provide the following constraints
\begin{align}
    N_1(t,r) &= -2 r N(r) \phi'(r) \phi_1(t,r) + b(r) \;, \\
    \delta_1'(t,r) &= -2 r \phi'(r) \phi_1'(t,r) \;, \\
    V_1'(t,r) &= -V'(r)\biggl( \delta_1(t,r) + \frac{\phi_1(t,r)}{F[\phi]} \frac{\delta F[\phi]}{\delta \phi} \biggr) + c(r) \;,
\end{align}
where $b(r)$ and $c(r)$ are free radial functions arising as the integration constants with respect to time. In the late time limit we expect the perturbation functions $\{N_1, \delta_1,\phi_1,V_1\}$ to vanish, corresponding to the unperturbed solution; this automatically evaluates the integration constants as $b(r)=c(r)=0$.
By performing a field redefinition $\Psi(t,r) = r \phi_1(t,r)$, the equations above  can be recast into a one-dimensional wave equation
\begin{equation}\label{eq:t_radial_eq}
    \biggl( \frac{\partial^2}{\partial r_*^2} - \frac{\partial^2}{\partial t^2} \biggr) \Psi(t,r_*) = V_\phi \Psi(t,r_*) \;,
\end{equation}
where the tortoise coordinate is again defined by $dr_*/dr = e^{\delta}/N$, and the potential reads 
\begin{align}\label{eq:pot_scal_radial}
    V_\phi &= \frac{e^{-2 \delta}N}{r^2}\biggl\{ 1-N-2r^2\phi'^2 -\frac{Q^2}{2 r^2}\biggl[ \frac{2}{F[\phi]}(1-2 r^2 \phi'^2) \nonumber \\
    & - \frac{2}{F^3[\phi]}\biggl(\frac{\delta F[\phi]}{\delta \phi}\biggr)^2 \nonumber \\
    & + \frac{1}{F^2[\phi]} \biggl(\frac{\delta^2 F[\phi]}{\delta \phi^2} + 4 r \phi' \frac{\delta F[\phi]}{\delta \phi}\biggr) \biggr] \biggr\} \;.
\end{align}
This expression is identical to the one in~\cite{Fernandes:2019rez}.
In the specific case given by Eq.~\eqref{coupling} the potential $V_\phi$ becomes
\begin{equation}
    V_\phi = \frac{N e^{-2 \delta}}{r^2} \biggl\{ 1- N - 2r^2 \phi'^2 - \frac{Q^2}{r^2 e^{\alpha \phi^2}}[1 + \alpha - 2(\alpha \phi - r \phi')^2] \biggr\} \;,
\end{equation}
or, in terms of the Misner-Sharp mass function $m(r)$,
\begin{align}
    V_\phi &= \frac{(r-2m)e^{-2\delta}}{r^3}\biggl\{ \frac{2m}{r} - 2r^2 \phi'^2 \nonumber \\
    & -\frac{Q^2}{r^2 e^{\alpha \phi^2}} [1 + \alpha - 2(\alpha \phi - r \phi')^2] \biggr\} \;.
\end{align}
It is straightforward to show that for $\alpha = \phi = \delta = 0$ and $m(r) = M - \frac{Q^2}{2 r}$ the latter reduces to the effective potential of linear and radial perturbations of a test scalar field in the RN background.

\begin{figure}[h]
  \centering
  \includegraphics[width=\columnwidth]{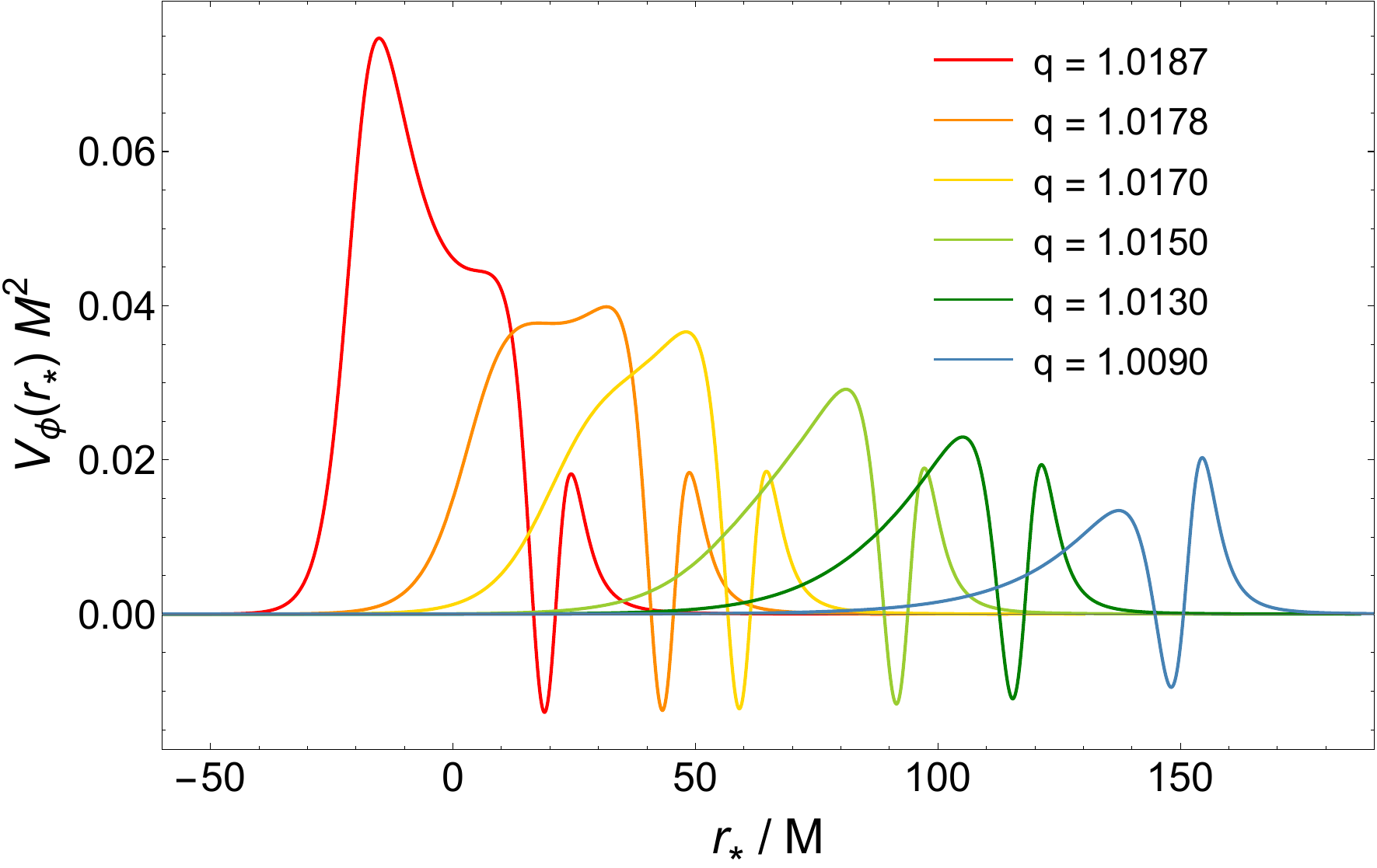}
  \includegraphics[width=\columnwidth]{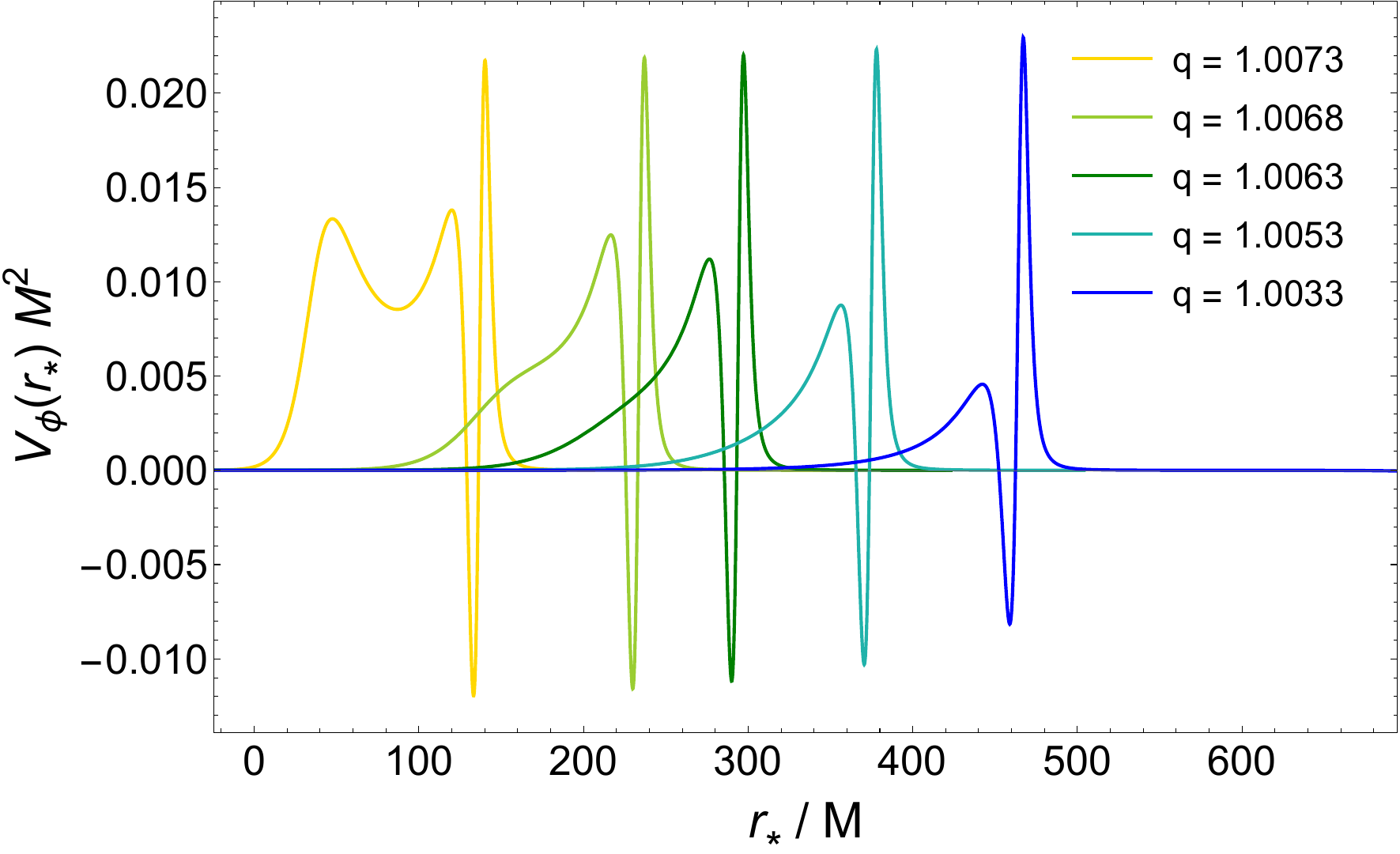}
  \caption{Effective potential $V_\phi$ for radial perturbations of scalarized BHs in terms of the tortoise coordinate $r_*$. We show the case $\alpha = 0.6$ in the top panel, and $\alpha = 0.5$ in the bottom panel. Different colors refer to different choices of $q = Q/M$.}\label{fig:Vphi}
\end{figure}

To study the QNMs we express the scalar field in terms of its Fourier transform, $\Psi(t,r) = \int d\omega e^{-i \omega t} \Tilde{\Psi}(\omega,r)$
and, dropping the tilde notation, we rewrite Eq.~\eqref{eq:t_radial_eq} as
\begin{equation}\label{eq:Schro_radial_eq}
    \biggl( \frac{d^2}{d r_*^2} + \omega^2 \biggr) \Psi = V_\phi \Psi \;.
\end{equation}
In the parameter space region where the hairy BH shows a stable photon sphere, the effective potential $V_{\phi}$ does not simply resemble that of a standard BH. Indeed, as shown in Fig.~\ref{fig:Vphi}, it can develop one or more cavities that would affect the associated QNMs.
Note that the potential can be negative in some regions, but overall
the area $\int_{-\infty}^{+\infty} dr_* V_\phi>0$, so there are no obvious instabilities. Indeed, as discussed later, we only found stable QNMs even when the potential has some negative regions.
Nevertheless, such a multi-peaked potential can support long-lived modes that, in the time domain, can leak through the potential barrier and appear as echoes in the ringdown signal. The computation of the QNMs and the time-domain analysis of radial perturbations will be discussed in Sec.~\ref{sec:time_QNMs}.

\subsection{Non-spherical perturbations}
In this section we will derive the linearized  equations governing non-spherical perturbations of the BH solution~\eqref{eq:bgBH_sol}. Let us consider the perturbed metric tensor $g_{\mu\nu} = \bar{g}_{\mu\nu} + h_{\mu\nu}$, where $\bar{g}_{\mu\nu}$ represents the background line element. In the same fashion we introduce the perturbed vector potential and scalar field as $A_{\mu} = \bar{A}_\mu + \delta A_{\mu}$ and  $\phi = \bar{\phi} + \delta \phi$, respectively. Since the background is spherically symmetric, we can decompose the field perturbations in terms of scalar, vector and tensor spherical harmonics with harmonic index $l$ and azimuthal number $m$. This procedure allows the splitting of the perturbations into  “axial'' (which acquire a factor $(-1)^{l+1}$ under parity inversion) and “polar'' (which acquire a factor $(-1)^l$). Since we are interested in the gravitational perturbations, in the following we will focus on the $l\geq2$ case\footnote{Note that for $l=1$ the only propagating degrees of freedom are the scalar and a polar electromagnetic mode.}. 

Let us expand the metric tensor perturbations in tensor spherical harmonics using the Regge-Wheeler gauge~\cite{PhysRev.108.1063}
\begin{align}
    h_{\mu\nu} &= h_{\mu\nu}^{A} + h_{\mu\nu}^{P} \;, \\
    h_{\mu\nu}^{A} &= \sum_{l,m} \int d\omega \, e^{-i \omega t} \,\times  \nonumber \\
    &\hspace{1.0cm}\begin{bmatrix}
        0 & 0 & -\frac{h_0(r) \partial_\varphi Y_l^m}{\sin \theta} & h_0(r) \sin \theta \partial_\theta Y_l^m \\[1ex]
        * & 0 & -\frac{h_1(r) \partial_\varphi Y_l^m}{\sin \theta} & h_1(r) \sin \theta \partial_\theta Y_l^m \\[2ex]
        * & * & 0 & 0 \\[1ex]
        * & * & * & 0
    \end{bmatrix} \;, \\
     h_{\mu\nu}^{P} &= \sum_{l,m} \int d\omega \, e^{-i \omega t} Y_l^m \,\times \nonumber \\
     &\hspace{-0.5cm}\begin{bmatrix}
        e^{-2 \delta(r)} N(r) H_0(r) & H_1(r) & 0 & 0 \\
        * & \frac{H_2(r)}{N(r)} & 0 & 0 \\
        * & * & r^2 K(r) & 0 \\
        * & * & * & r^2 \sin^2 \theta K(r)
    \end{bmatrix} \;,
\end{align}
where the asterisk ``$*$"  here denotes symmetrization, $Y_l^m = Y_l^m (\theta, \varphi)$ are the scalar spherical harmonics, and $A$ and $P$ stand for “axial'' and “polar'', respectively. The perturbation of the vector potential is decomposed similarly
\begin{align}
    \delta A_\mu &= \delta A_\mu^A + \delta A_\mu^P \;,\\
     \delta A_\mu^A &= \sum_{l,m} \int d \omega \, e^{-i \omega t} \,\times \nonumber \\
     &\hspace{1.0cm}\biggl( 0, 0, -\frac{u_4(r)\partial_\varphi Y_l^m}{\sin \theta}, u_4(r) \sin \theta \partial_\theta Y_l^m \biggr)\;,\\
      \delta A_\mu^P &=  \sum_{l,m} \int d \omega \, e^{-i \omega t} \biggl( \frac{u_1(r) Y_l^m}{r}, \frac{u_2(r) Y_l^m}{r N(r)}, 0, 0 \biggr)\;,
 \end{align}
where we have fixed the gauge of the vector potential by setting the angular components of $\delta A_\mu^P$ to zero. However, for the next derivation it is rather useful to consider the perturbations of the Maxwell tensor $F_{\mu\nu}$ as dynamical variable
\begin{align}
    \delta F_{\mu\nu} &= \delta F_{\mu\nu}^A + \delta F_{\mu\nu}^B \;,\\
    \delta F_{\mu\nu}^A &= \sum_{l,m} \int d\omega \, e^{-i\omega t} \,\times \nonumber \\
    &\hspace{0.5cm}\begin{bmatrix}
        0 & 0 & -\frac{i \omega u_4(r) \partial_\varphi Y_l^m}{\sin \theta} & i \omega u_4(r) \sin \theta \partial_\theta Y_l^m \\[1ex]
        \star & 0 & \frac{u'_4(r) \partial_\varphi Y_l^m}{\sin \theta} & -u'_4(r) \sin \theta \partial_\theta Y_l^m \\[1ex]
        \star & \star & 0 & l(l+1) u_4(r) \sin \theta Y_l^m \\[1ex]
        \star & \star & \star & 0
    \end{bmatrix} \; ,\\
    \delta F_{\mu\nu}^P &= \sum_{l,m} \int d\omega \, e^{-i \omega t} \,\times \nonumber \\
    &\hspace{0.5cm}\begin{bmatrix}
        0 & f_{01}(r)Y_l^m & f_{02}(r) \partial_\theta Y_l^m & f_{02}(r) \partial_\varphi Y_l^m \\[1ex]
        \star & 0 & f_{12}(r) \partial_\theta Y_l^m & f_{12}(r) \partial_\theta Y_l^m \\[1ex]
        \star & \star & 0 & 0 \\[1ex]
        \star & \star & \star & 0 
    \end{bmatrix} \; ,
\end{align}
where the star ``$\star$" denotes antisymmetrization. The functions $f_{01}$, $f_{02}$ and $f_{12}$ can be written in terms of $u_1$ and $u_2$ as follows
\begin{align}
    f_{01}(r) =& \frac{i \omega r u_2(r) + N(r)(r u_1'(r) - u_1(r))}{r^2 N(r)} \;, \\
    f_{02}(r) =& \frac{u_1(r)}{r} \;, \\
    f_{12}(r) =& \frac{u_2(r)}{r N(r)} \;,
\end{align}
and are related by the following equation
\begin{equation}\label{eq:bianchi}
    f_{01}(r) = i \omega f_{12}(r) + f_{02}'(r) \;.
\end{equation}
Finally, the scalar field perturbation is decomposed as
\begin{equation}
    \delta \phi = \sum_{l,m} \int d\omega \, e^{-i \omega t} z(r) Y_l^m \;.
\end{equation}

The full derivation of the linearly perturbed equations is outlined in Appendix~\ref{app:App_der_pert}.
\subsubsection{Axial sector}
The axial sector, which contains only perturbations of the Maxwell tensor and of the metric, can be written as a system of two coupled second order differential equations 
\begin{align}
    \biggl( \frac{d^2}{d r_*^2} + \omega^2 \biggr) U(r) =& V_{UU} U(r) + V_{UH} H(r) \;, \label{eq:axial_U}\\ 
    \biggl( \frac{d^2}{d r_*^2} + \omega^2 \biggr) H(r) =& V_{HU} U(r) + V_{HH} H(r) \label{eq:axial_H} \;,
\end{align}
where $U(r)$ and $H(r)$, defined in Eq.~\eqref{eq:U_H_defs}, are functions of $u_4(r)$ and $h_1(r)$, respectively. The potentials are given by
\begin{align} \nonumber
    V_{UU} =& \frac{r -2 m}{r^3 e^{2 \delta}} \biggl\{ \Lambda + 2 \\ \nonumber
    & \hspace{0.5cm}+ \frac{Q^2}{4 r^2 F^3[\phi]}\biggl[ 16 \, F^2[\phi] - \biggl( \frac{\delta F[\phi]}{\delta \phi} \biggr)^2 \biggr] \\ \nonumber
    & \hspace{0.5cm} + \frac{(r-2m) \phi'}{4 F[\phi]}\biggl[ 
2 r \phi' \frac{\delta^2 F[\phi]}{\delta \phi^2}  \\
& \hspace{1.5cm}-4 \frac{\delta F[\phi]}{\delta \phi} - \frac{r \phi'}{F[\phi]} \biggl( \frac{\delta F[\phi]}{\delta \phi} \biggr)^2 \biggr] \biggr\} \; ,\\
    V_{UH} =& V_{HU} = \frac{r -2 m}{r^3 e^{2 \delta}} \biggl(\frac{2 \sqrt{\Lambda}}{\sqrt{F[\phi]}} \frac{Q}{r} \biggr) \; , \\ \nonumber 
    V_{HH} =& \frac{r -2 m}{r^3 e^{2 \delta}} \biggl[ \Lambda - \frac{2m}{r} + 2m' \\
    &\hspace{2.0cm}+ 2\biggl( 1 - \frac{2m}{r} \biggr) + (r-2m)\delta' \biggr] \; ,
\end{align}
with $\Lambda \equiv (l+2)(l-1)$. 
As shown in \cite{Jansen:2019wag}, a system of this type can be decoupled via an $r-$independent transformation if the potentials satisfy the following requirement
\begin{align}
    \frac{V_{UU} - V_{HH}}{V_{UH}} = const. 
\end{align}
While this is the case in the RN limit, as expected, we verified numerically that this requirement is not fulfilled in general for scalarized BHs.
If we consider the EMS model in Eq.~\eqref{coupling}, the potentials simplify to
\begin{align}
    V_{UU} =& \frac{r -2 m}{r^3 e^{2 \delta}} \biggl[ \Lambda^2 +2  - \frac{e^{-\alpha \phi^2} Q^2 (\alpha^2 \phi^2 -4)}{r^2} \nonumber \\ 
    & \hspace{0.5cm}+ \alpha \phi'(r- 2 m)(r \phi' -2 \phi + r \alpha \phi' \phi^2)  \biggr] \; ,\\
    V_{UH} =& V_{HU} = \frac{r -2 m}{r^3 e^{2 \delta}} \biggl(2 \sqrt{\Lambda} e^{-\frac12 \alpha \phi^2 } \frac{Q}{r} \biggr) \; , \\
    V_{HH} =& \frac{r -2 m}{r^3 e^{2 \delta}} \biggl[ \Lambda^2 - \frac{2m}{r} + 2m' \nonumber \\
    & \hspace{1.5cm}+ 2\biggl( 1 - \frac{2m}{r} \biggr) + (r-2m)\delta' \biggr] \; .
\end{align}

\subsubsection{Polar sector}
The polar group (containing scalar, electromagnetic and metric perturbations) is described by a system of five first-order differential equations 
\begin{equation}\label{eq:polar_f12_gen}
     f_{12}'(r) = \biggl( \delta' - \frac{N'}{N} - \frac{1}{F[\phi]} \frac{\delta F[\phi]}{\delta \phi} \phi' \biggr) f_{12}(r) - \frac{i \omega e^{2 \delta}}{N^2} f_{02}(r) \; ,
\end{equation}
\begin{align}\label{eq:polar_H1_gen}
    H_1'(r) &= \biggl( \delta' - \frac{N'}{N} \biggr) H_1(r) \nonumber \\
    & - \frac{i \omega}{N} [H_0(r) + K(r)] - 4 F[\phi] V' f_{12}(r) \; ,
\end{align}
\begin{align}\label{eq:polar_f02_gen}
    f_{02}'(r) &= \biggl[ \frac{ l(l+1) N e^{-2 \delta}}{r^2 \omega} - \omega \biggr] i f_{12}(r) \nonumber \\
    & - V' K(r) - \frac{V'}{F[\phi]} \frac{\delta F[\phi]}{\delta \phi} \, z(r) \; ,
\end{align}
\begin{align}\label{eq:polar_K_gen}
    K'(r) =& \frac{H_0(r)}{r} - \biggl( \frac{1}{r} + \delta' -\frac{N'}{2 N} \biggr) K(r) - 2 \phi' z(r) \notag \\
    &+ \frac{i}{2 r^2 \omega} [l(l+1) -2 + 2 r N' + 2 e^{2 \delta} r^2 V'^2 F[\phi] \nonumber \\
    & +2 N (1 + r^2 \phi'^2)]H_1(r) \; ,
\end{align}
\begin{align}\label{eq:polar_H0_gen}
     H_0'(r) =& - \frac{4 V' e^{2 \delta}F[\phi]}{N} f_{02}(r) + \biggl( \frac{N'}{2N} - \delta' -\frac{1}{r} \biggr) K(r) \nonumber \\
     & + \biggl( \frac{1}{r} + 2 \delta' -\frac{N'}{N} \biggr) H_0(r) + 2 \phi' z(r) \notag \\ 
    & - \biggl\{ \frac{i \omega e^{2 \delta}}{N} - \frac{i}{2 r^2 \omega}\biggl[l(l+1) - 2 + 2 r N' \nonumber \\
    & + 2 r^2 V'^2 e^{2\delta} F[\phi] + 2N + 2 r^2 N \phi'^2\biggr] \biggr\}H_1(r) \; ,
\end{align}
and a second-order one,
\begin{align}\label{eq:polar_z_gen}
        z''(r) =& -\frac{i}{r^2 \omega} l(l+1) V' \frac{\delta F[\phi]}{\delta \phi} f_{12}(r)  \nonumber \\
        & + \frac{1}{N} V'^2 e^{2 \delta} \frac{\delta F[\phi]}{\delta \phi} K(r) - \frac{4}{N} \phi' V' e^{2 \delta} F[\phi] f_{02}(r) \notag \\
    &- \frac{1}{2 N}[V'^2 e^{2 \delta} \frac{\delta F[\phi]}{\delta \phi} + 2(N' - 2 N \delta')\phi']H_0(r) \nonumber \\
    &+ \biggl( \delta' - \frac{2}{r} - \frac{N'}{N} \biggr) z'(r) \notag \\ 
    &+ \bigg\{ 4 \phi'^2 + \frac{1}{N}\biggl[ \frac{l(l+1)}{r^2} - e^{2 \delta} \omega^2 \nonumber \\
    & - \frac{e^{2 \delta} V'^2}{2 F[\phi]}\biggl( F[\phi] \frac{\delta^2 F[\phi]}{\delta \phi^2} -2 \biggl( \frac{\delta F[\phi]}{\delta \phi} \biggr)^2\biggr) \biggr]\biggr\} z(r) \;.
\end{align}

\section{QNMs and time-domain linear response}\label{sec:time_QNMs}

In this section we present the results of the perturbative study of the EMS BH. Sections~\ref{sec:robin-iv-a} and~\ref{sec:robin-iv-b} outline the methods used for the QNM approach and time domain approach, respectively. Sections~\ref{sec:QNM_linear_radial} and~\ref{sec:robin-iv-d} discuss the results of the QNM and time domain approach for the purely radial case. Finally, section~\ref{sec:robin-iv-e} presents the numerical results for the time domain simulations of axial modes.

\subsection{Frequecy domain method}
\label{sec:robin-iv-a}

We compute the QNMs obtained by solving the equations previously derived as an eigenvalue problem, imposing the usual purely outgoing boundary conditions at infinity and purely ingoing boundary conditions near the BH horizon.
For a given value of $l$, this selects a discrete set of complex frequencies $\omega=\omega_R+i\omega_I$ which can be labeled by the overtone index $n=0,1,2,3$, with $n=0$ being the fundamental mode (the one with smaller absolute value of the imaginary part) and $n=1,2,3,...$ being the overtones.
The QNM frequencies are evaluated numerically using a direct integration shooting method (see, e.g., \cite{Ferrari:2007rc,Rosa:2011my,Pani:2012bp,Pani:2013pma}).

Close to the horizon the solution of the perturbation equations behaves like a superposition of ingoing and outgoing waves, but BH boundary conditions require only the former. The perturbation (or an array of perturbations) can be expanded near the horizon as follows
\begin{equation}
    \Psi \sim (r-r_H)^\lambda \sum_{i=0}^\infty \Psi_{H,i} (r-r_H)^i \,,
\end{equation}
and solving the perturbation equations as a Frobenius series we get the exponent $\lambda$ and the coefficients of the expansions $\Psi_{H,i \ge 1}$ in terms of $\Psi_{H,0}$. In the case of a single equation (as in the case of radial perturbations) we can simply set $\Psi_{H,0} = 1$, thanks to the linearity of the equation. For a coupled system of $N$ second order ODEs we choose a basis of $N-$dimensional unit vectors. With these boundary conditions for the inner boundary we can integrate the equations from a point near the horizon $\bar{r}_H = r_H (1+\epsilon)$, with $\epsilon \ll 1$ (to avoid numerical divergences), to a large enough numerical value that represents infinity.

Similarly, at infinity the general solution of the perturbation equations is expected to be a linear combination of ingoing and outgoing waves as follows
\begin{equation}
    \Psi \sim B e^{-i \omega r} r^{\bar{\lambda}} + C e^{i \omega r} r^{-\bar{\lambda}} \,,
\end{equation}
where the $B$ and $C$ are constants depending on the complex frequency $\omega$, while the exponent $\bar{\lambda}$ can be derived by solving the equations expanded asymptotically at infinity. As anticipated, the QNMs can be found by imposing the proper boundary conditions at infinity. Therefore we allow for outgoing waves only, i.e. we require the constant $B$ to be zero. This applies straightforwardly to the case of radial perturbations, where we have a single master equation~\eqref{eq:Schro_radial_eq}. In the case of a system of $N$ coupled second order ODEs with $N$ perturbations (like Eqs.~\eqref{eq:axial_U} and~\eqref{eq:axial_H} with $N=2$) we need to construct a matrix by using an orthogonal basis for the $N-$dimensional space and then request its determinant to be zero (for more details on this procedure see \cite{Pani:2012bp,Pani:2013pma}).
Being the constant $B$ (or the determinant) a complex number, we impose the null condition via a shooting procedure in the complex $\omega$ plane. 
The calculations described above are carried out using \textsc{Mathematica}, employing high numerical precision and high-order series expansions near both boundaries for improved accuracy.

\subsection{Time domain method} 
\label{sec:robin-iv-b}

Furthermore, we evolve the perturbative equations of motion for scalar and axial modes, in the time domain, with a purpose built \CC{} code implementing the method of lines. This evolution code discretises space into a uniform grid with gridpoint coordinates given by $x_i = x_0 + i \, \Delta x$, where $\Delta x$ is the grid spacing. On this grid, any desired field $f(t,x)$ is also discretised into a set of $f_i(t)$ where $f_i(t) = f(t,x_i)$. In contrast to finding the QNMs, we require a wave equation with time derivatives present rather than the eigenfrequencies $\omega$. For the scalar perturbation this is given by Eq.~\eqref{eq:t_radial_eq}; this equation must be further decomposed into two first order ODEs in time as follows,
\begin{align} \label{eq:robin-scalar-eom}
\frac{\partial}{\partial t}\dot{\Psi}(t,r_*) &= \frac{\partial^2}{\partial r_*^2} \Psi(t,r_*) - V_\phi(r_*) \Psi(t,r_*), \\
\frac{\partial}{\partial t}\Psi(t,r_*) &= \dot{\Psi}(t,r_*).
\end{align}
Both $\Psi$ and $\dot{\Psi}$ are used as evolution variables. Note that we require input values of $V_\phi(x_i)$, which encodes the background spacetime, along with suitable initial data (discussed later in Eqs.~\ref{eq:time-domain-intial-pert1} $\&$ \ref{eq:time-domain-intial-pert2}).  From now on we will set $r_* = x$ for simplicity. 

On a grid point $x_i$, spatial derivatives of $\Psi_i(t)$ are evaluated using 4th order derivative stencils of the form
\begin{align}
    \frac{\partial^n}{\partial x^n} \Psi_i(t) = \frac{1}{(\Delta x)^n}\sum_{j=i-2}^{i+2} \Delta^{(n)}_{j}\Psi_j.
\end{align}
For example, in the case of a symmetric second derivative $\Delta_j^{(2)} = \{-1/12, 4/3,-5/2 ,4/3,-1/12\}$, fully written as
\begin{align}
    \frac{\partial^2}{\partial x^2}\Psi_i(t) =& \frac{1}{12\Delta x^2} \Big[-\Psi_{i-2}(t)  + 16 \Psi_{i-1}(t) - 30 \Psi_i(t) \nonumber\\ &\quad\quad\quad\,\,\,\,\,\,\,\, + 16 \Psi_{i+1}(t) -\Psi_{i+2}(t)\Big];
\end{align}
effectively spatial derivatives have been transformed into linear functions of field values depending on a small neighbourhood around $x_i$. 

Finally, at each gridpoint, the time evolution of $\{\Psi_i,\dot{\Psi}_i\}$ is done using the standard Runge-Kutta 4-th order method. This combination of spatial derivative stencils and Runge-Kutta time integration is known as the method of lines.

The time evolution of the axial modes is almost the same as for the scalar mode. The main change is the different equations of motion, which can be recovered from Eqs.~\eqref{eq:axial_U} and \eqref{eq:axial_H},
\begin{align} 
\frac{\partial}{\partial t}\dot{U}(t,x) &= \frac{\partial^2}{\partial x^2} U(t,x) - V_{UU}U(t,x) - V_{UH} H(t,x), \\
\frac{\partial}{\partial t}\dot{H}(t,x) &= \frac{\partial^2}{\partial x^2} H(t,x) - V_{UH}U(t,x) - V_{HH} H(t,x), \\
\frac{\partial}{\partial t}U(t,x) &= \dot{U}{\Psi}(t,x), \\
\frac{\partial}{\partial t}H(t,x) &= \dot{H}{\Psi}(t,x).
\end{align}
The set of evolution variables $\{U,\dot{U}, H, \dot{H}\}$ can be evolved in time using the method of lines, as before.

The boundary conditions used are of Sommerfeld type, also known as ``outgoing wave" type. The potentials $\{V_\phi,V_{UU}, V_{UH}, V_{HH} \}$ are assumed to be negligibly small at the boundaries; violating this assumption leads to undesirable reflections from the boundary conditions which can poison the delicate echo signal we seek to measure. The Sommerfeld boundaries have been imposed in the code by requiring that, for every field variable $\psi \in \{\Psi,\dot{\Psi}\}$ or $\{U,\dot{U}, H, \dot{H}\}$,
\begin{align}
    \frac{\partial}{\partial t}\psi(t,x) \pm \frac{\partial}{\partial x}\psi(t,x) = 0.
\end{align}
These equations are designed to let the fields $\psi_n$ propagate past the left/right hand side of the grid by picking the $+/-$ sign. Note that as there are no numerical gridpoints on the left/right side of the left/right boundary we must use one-sided derivative stencils.

\subsection{QNMs of linear radial perturbations} \label{sec:QNM_linear_radial}
\begin{figure*}[th]
  \centering
  \includegraphics[width=1\textwidth]{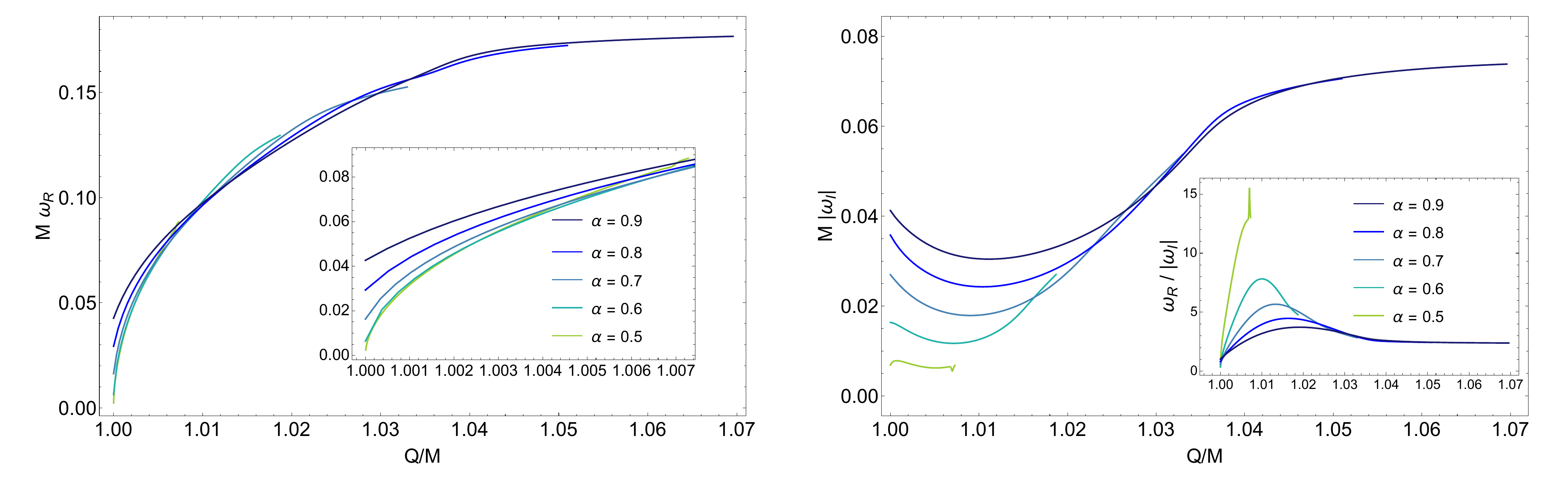}
  \caption{Fundamental QNMs for linear radial perturbations as a function of the charge-to-mass ratio $q=Q/M$ for different values of the coupling constant $\alpha$. Note that we refer to the fundamental mode as the one with the least damping for $Q / M \approx 1$, but increasing $Q / M$ this mode ceases to be the least damped due to a crossing in the imaginary part of the frequency with other modes. The peak present in the imaginary part of the mode for $\alpha = 0.5$ and shown in the inset of the right panel represents a genuine physical behavior, verified by increasing the resolution.
  }
  \label{fig:QNMs_radial}
\end{figure*}

\begin{figure*}[th]
  \centering
  \includegraphics[width=1\textwidth]{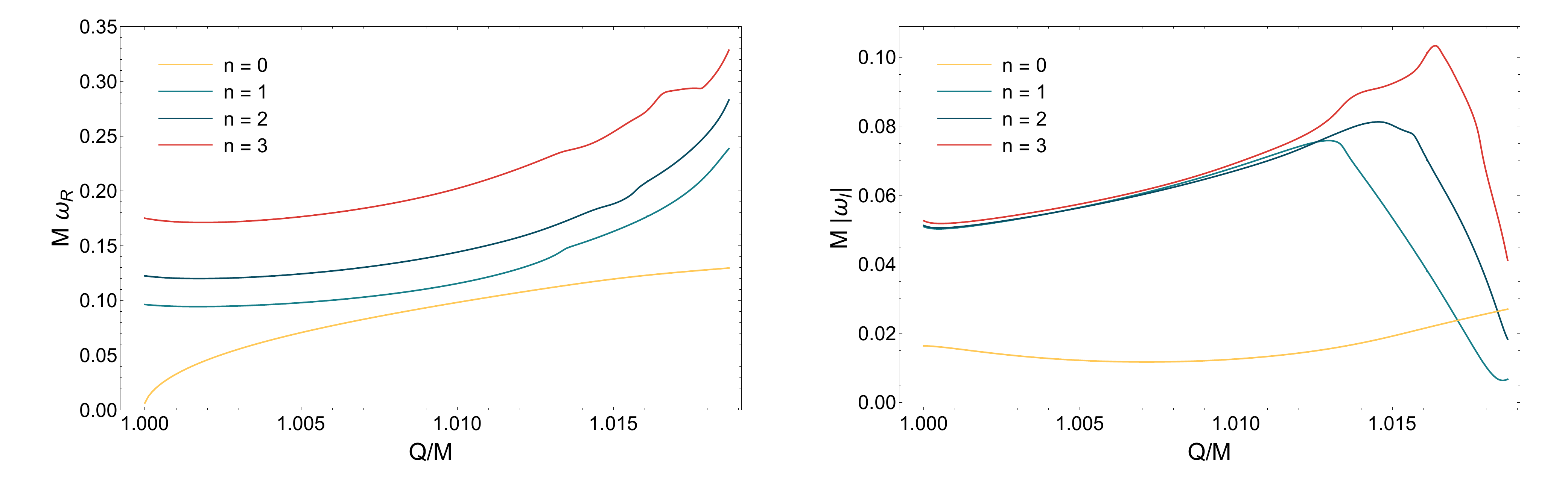}
  \includegraphics[width=1\textwidth]{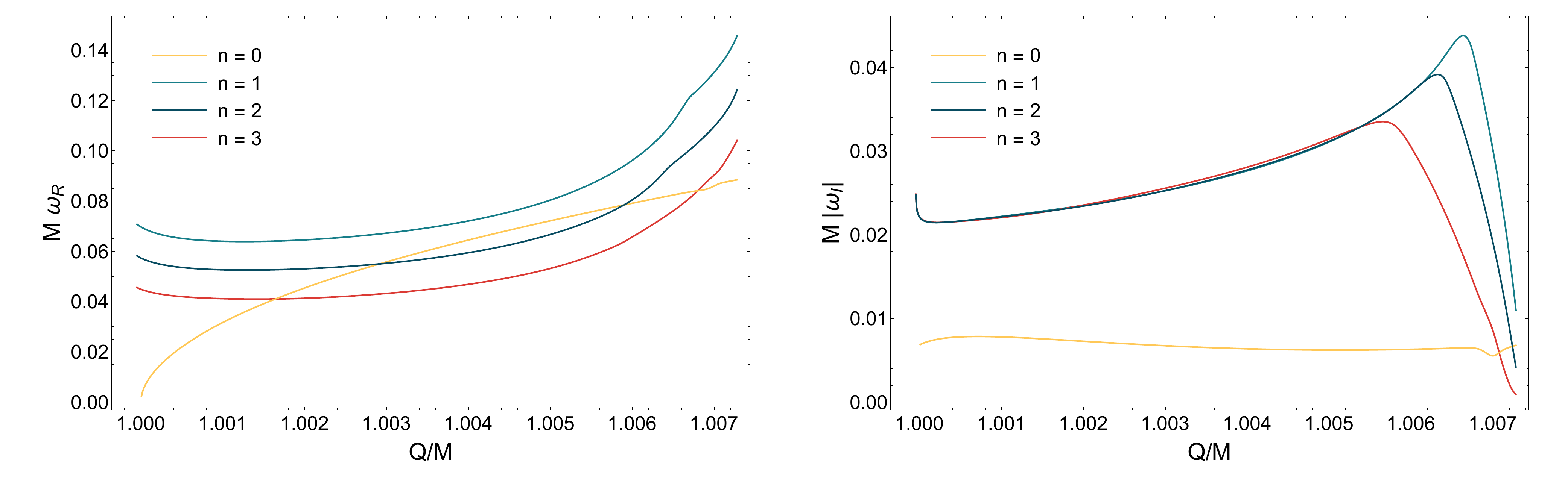}
  \caption{Fundamental QNM and first three overtones for linear and spherical perturbations as a function of the charge-to-mass ratio $q=Q/M$ for $\alpha = 0.6$ (top panels) and $\alpha = 0.5$ (bottom panels).}\label{fig:QNMs_radial2}
\end{figure*}

The QNMs of linear radial perturbations of the scalarized BH are shown in Figs.~\ref{fig:QNMs_radial} and \ref{fig:QNMs_radial2} as functions of the charge-to-mass ratio of the background. We focus on the regime $Q/M\gtrsim 1$, which is the regime where the effective potential develops multiple peaks.
Figure~\ref{fig:QNMs_radial} focuses on the fundamental mode  for different values of the coupling constant $\alpha$. Note that the real part is only mildly sensitive to $\alpha$ while the imaginary part displays a stronger dependence especially near $Q/M\approx 1$.
As expected from the behavior of the effective potential, the quality factor $\omega_R/|\omega_I|$ increases for smaller values of $\alpha$ and when $Q/M\approx 1$ (see inset in the right panel). 
For $\alpha=0.5$, the quality factor can be as high as $\approx 15$. As a comparison, the quality factor for a Schwarzschild BH is almost a factor four smaller.

In Fig.~\ref{fig:QNMs_radial2} we show also the first few overtones for $\alpha=0.6$ and $\alpha=0.5$, since these will be used to fit the time domain signal, as discussed in the next section.

\begin{figure*}[th]
  \centering
  \includegraphics[width=1\textwidth]{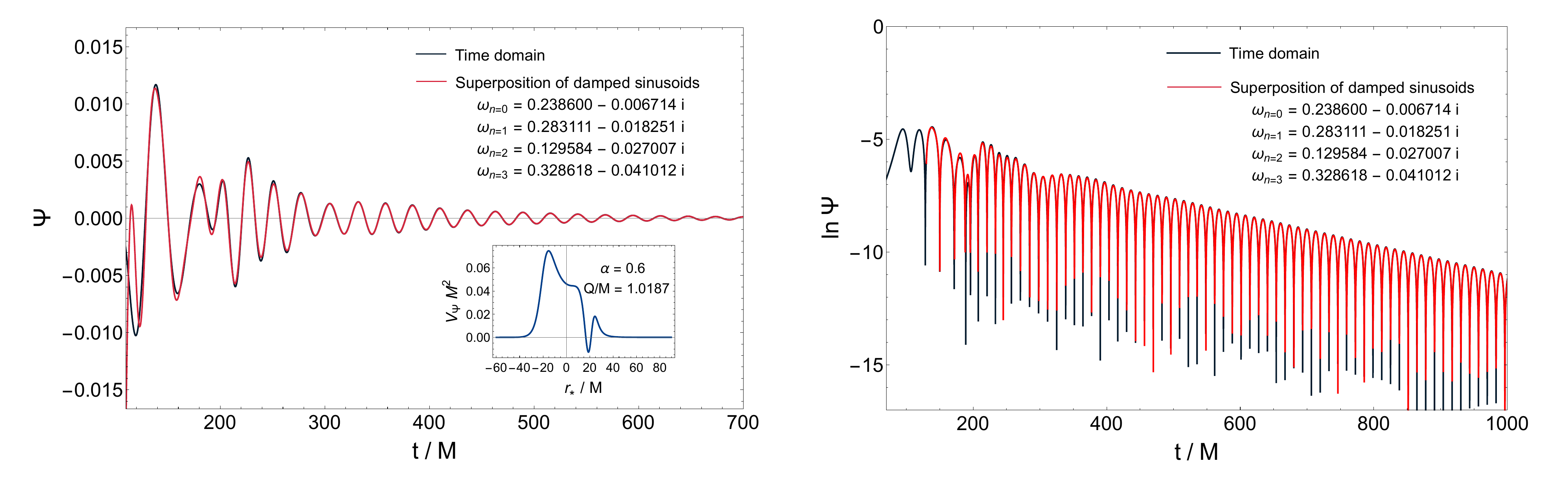}
  \includegraphics[width=1\textwidth]{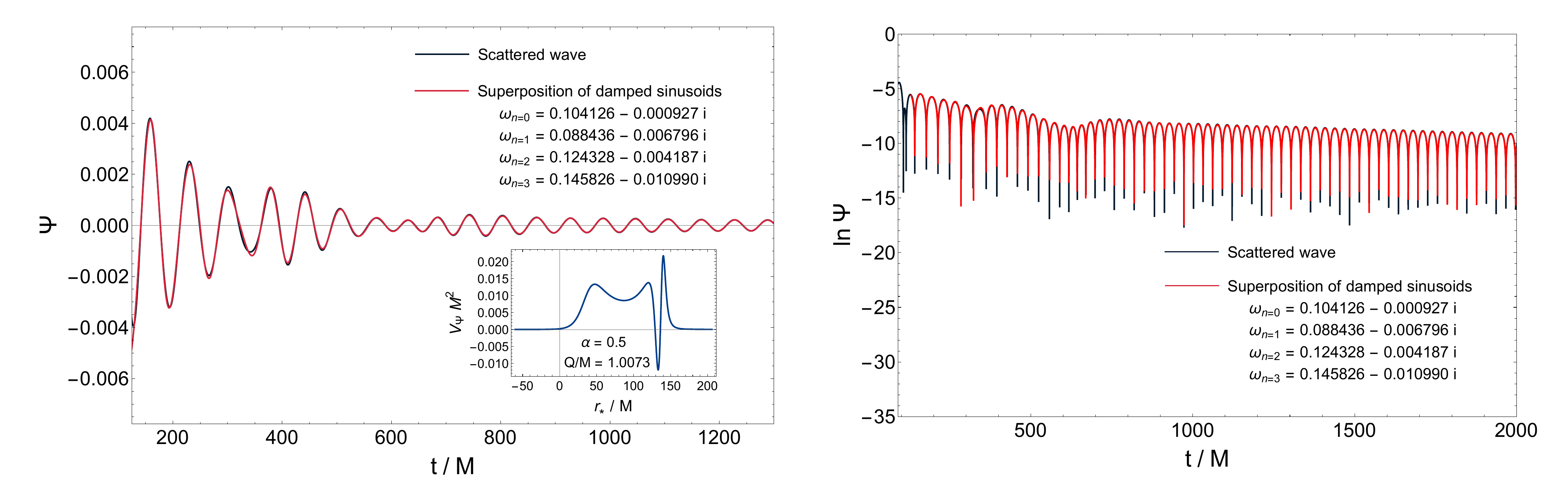}
  \caption{Comparison between the scattered wave in the time domain and a superposition of four damped sinusoids with the fundamental QNM and the first three overtones for $\alpha = 0.6$, $Q/M = 1.0187$ (top panels) and for $\alpha = 0.5$, $Q/M = 1.0073$ (bottom panels). The QNM used for the fit are the same as those shown in Fig.~\ref{fig:QNMs_radial2} and are provided in the plots. The inset shows the associated effective potential of the scalar perturbation equation. The right panel show the signal in a logarithmic scale, to better appreciate the echoes.}
  \label{fig:TD_linear}
\end{figure*}

\subsection{Time-domain linear spherical perturbations}
\label{sec:robin-iv-d}
We now move to analyze the linear radial perturbations in the time domain.
The initial data for the time-domain code consists of a Gaussian wave-packet (denoted by $\psi$) boosted towards the BH from large radius $x_0$, where the effective potential is small. The wave-packet propagates inwards as a result of the ansatz,
\begin{align} \label{eq:time-domain-intial-pert1}
    \psi(x,t) &= Ae^{\frac{(x-x_0+t)^2}{2\sigma^2}},\\
    \label{eq:time-domain-intial-pert2}
    \dot{\psi}(x,t) &= \frac{A(x-x_0 +t)}{\sigma^2}e^{\frac{(x-x_0+t)^2}{2\sigma^2}},
\end{align}
and is then incident on the outside of the potential well. Note that $A$ and $\sigma$ are constants representing the initial amplitude and standard deviation of the initial wave-packet. Most of the initial wave is reflected due to the steep nature or the potential well, but a fraction of the energy passes the barrier and excites the cavity. The cavity supports a quasi-long lived and quasi-periodic excitation that slowly decays via radiation being emitted from each side of the potential well. The signal emitted to large radius is then recorded at a fixed radius.

A representative summary of our results is presented in Fig.~\ref{fig:TD_linear}, where we show the time-domain response of a wave scattered off the BH for selected values of $q$ and $\alpha$.
As shown in the insets, in this cases the potential displays multiple peaks, which can trap radiation and give rise to a complex ringdown signal. While the signal at late times is governed by the (relatively long-lived) fundamental QNM, initially one can clearly notice repeated modes that are reminiscent of echoes~\cite{Cardoso:2016rao,Cardoso:2016oxy,Cardoso:2017cqb}.
Indeed, these repeated modes are due to perturbations being trapped in the potential cavity. However, since the latter is not particularly wide and deep in this model, the trapping is inefficient and repeated echoes decay fast in time, being indeed more appreciable in a logarithmic scale (see right panels of Fig.~\ref{fig:TD_linear}.

Despite the complex time-domain signal due to cavity effects, it is interesting that a superposition of the first few QNMs fits the entire signal quite accurately. In Fig.~\ref{fig:TD_linear} we compare the signal with a fit 
\begin{equation}
   \Psi\sim \sum_{n=0}^3 A_n e^{i(\omega_R^{(n)} t+\phi_n)}e^{-t/\tau_n} \,,\label{superposition}
\end{equation}
where $A_n$ and $\phi_n$ are the amplitude and phase of the $n$-th tone and $\tau_n=1/|\omega_I^{(n)}|$ is the damping time. This superposition well reproduces also the nontrivial behavior at early times.

\subsection{Ringdown of non-spherical perturbations}
\label{sec:robin-iv-e}

For completeness, we now look at non-spherical perturbations, focusing on the axial sector; this is easier to solve numerically compared to the polar one.

\begin{figure*}[th]
  \centering
  \includegraphics[width=1\textwidth]{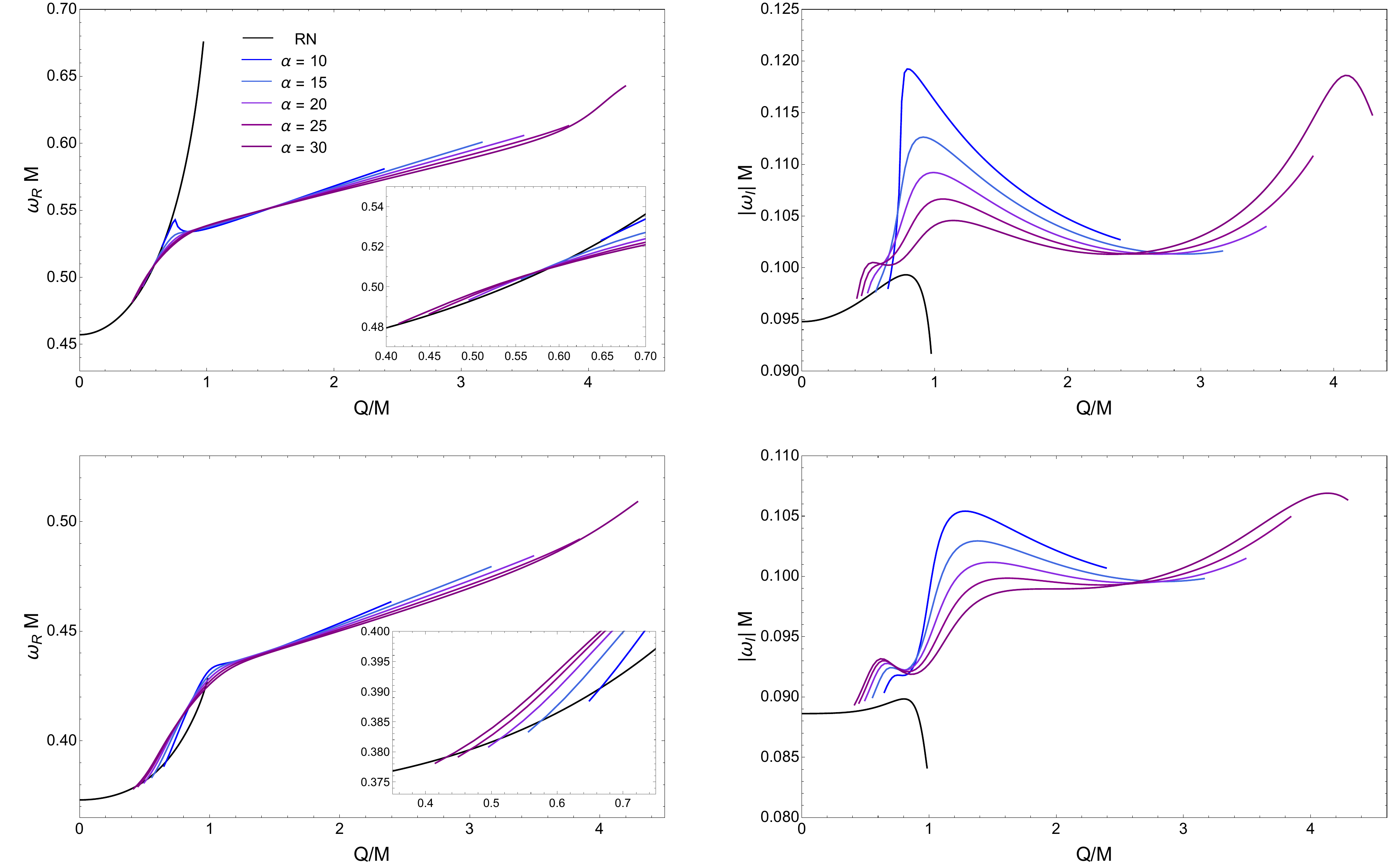}
  \caption{Real (left panels) and imaginary (right panels) frequencies of the $l=2$ QNMs for EM-led (top panels) and gravitational-led (bottom panels) modes as a function of the charge-to-mass ratio $q=Q/M$ for different values of the coupling constant $\alpha$. The limit of $\alpha = 0$ represents the RN QNMs shown in black.}\label{fig:QNM_nonradial}
\end{figure*}

In Fig.~\ref{fig:QNM_nonradial} we show the QNMs as a function of $q$ for different values of $\alpha$, comparing them with the RN case, which is also a static solution of the theory. In this case we consider higher values of $\alpha$, since the small-$\alpha$ region is difficult to study numerically, at least with our shooting method. 

We computed the QNMs frequencies across the full parameter space of the scalarized BHs, considering different values of the coupling constant $\alpha$. As can be seen from the insets of Fig.~\ref{fig:QNM_nonradial}, for the minimum values of $q$, the QNMs of the BHs with scalar hair are continuously connected to the QNMs of the RN BH. Since the Eqs. \eqref{eq:axial_U} and \eqref{eq:axial_H} are coupled, the QNMs are solutions to both equations and are not induced by a single perturbation. Nevertheless, we refer to modes as EM-led (or gravitational-led) if they reduce to those induced by electromagnetic (or gravitational) perturbations in the RN case. 
Focusing on the real part of the frequencies, for moderately low values of $q$ the differences with the RN case are small, particularly for the gravitational-led modes. The differences become more pronounced for larger values of $q$, especially for $q > 1$ where the RN solutions ceases to exist. As already discussed in Sec.~\ref{sec:QNM_linear_radial} for the radial perturbations, the real part is only weakly dependent on $\alpha$. The role of $\alpha$ is much more evident for the imaginary part, where notable differences arise not only compared to the RN case but also among different values of the coupling constant.

\begin{figure}
    \centering
    \includegraphics[width=\columnwidth]{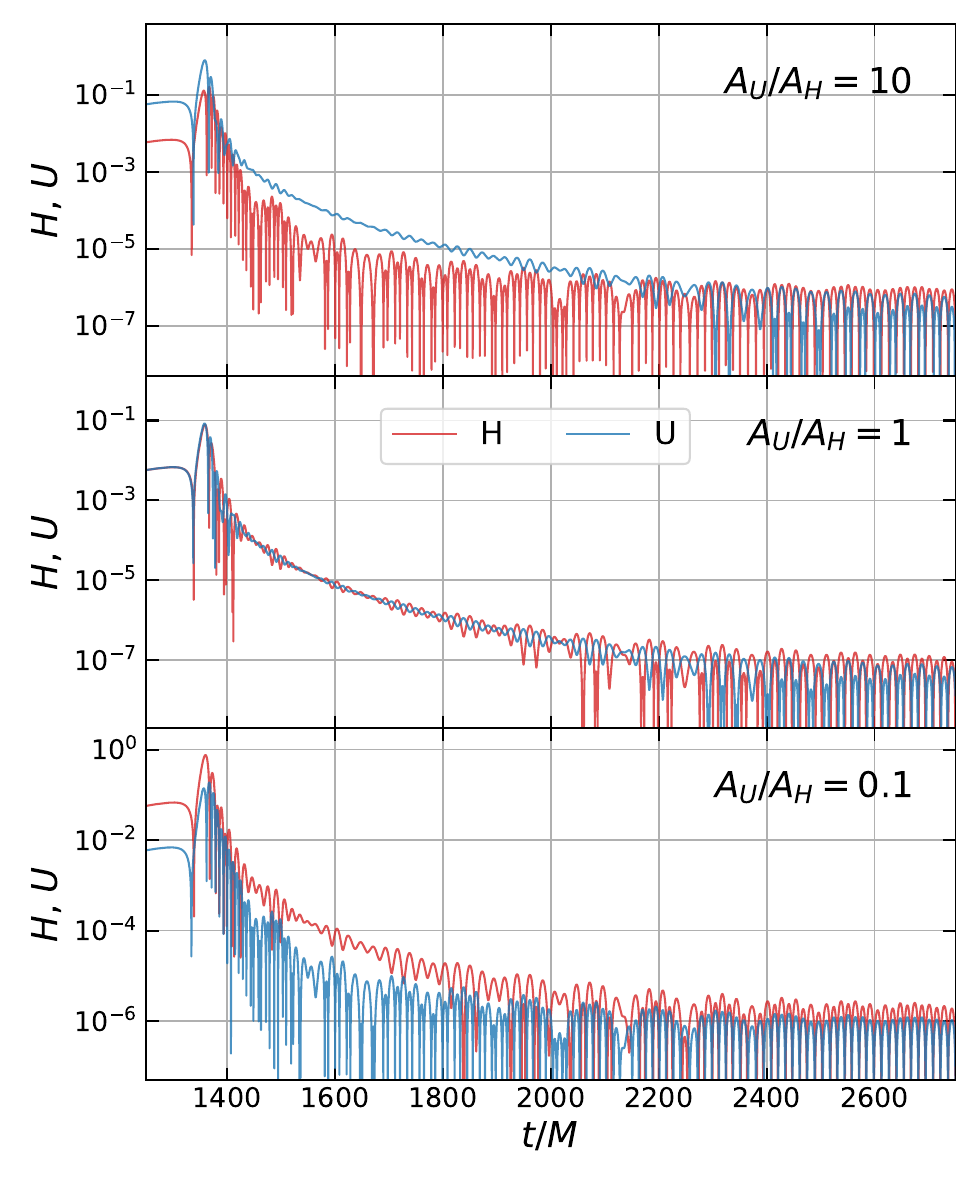}
    \caption{
    Time-domain response of a hairy BH for $\alpha=0.6$, $Q=1.0187 M$ and for $l=2$ axial perturbations. The panels show different values of $A_U/A_H$, where $A_U$ and $A_H$ are the amplitudes of the EM-led and gravitational-led perturbations, respectively. Echoes appear also in this case and their pattern depends on $A_U/A_H$, due to the coupling between axial EM and gravitational perturbations.}
    \label{fig:axialtime}
\end{figure}

Finally, for completeness in Fig.~\ref{fig:axialtime} we show the time-domain response of a hairy BH in the axial sector for $l=2$ and different values of the amplitudes of the initial wavepackets. From top to bottom, the three panels show the cases $A_U/A_H=\{10,1,0.1\}$, where $A_U$ and $A_H$ are the amplitudes of the EM-led and gravitational-led perturbations, respectively. Overall, we observe a behavior that is qualitatively similar to the radial scalar case. However, because EM-led and gravitational-led perturbations are coupled to each other, the echo pattern depends on $A_U/A_H$ and is in general more complex, since the cavity mode of one field can leak into the other.
In particular, the echo pattern is more evident when the initial data are dominated by either of the two fields. We expect this to be due to the role of the mixed potential $V_{UH}$.

\section{Nonlinear radial dynamics}

We will now discuss the fully-nonlinear numerical simulations of the spherical collapse of wave packets of the field $\phi$ onto scalarized BHs. 
Our main goal is to assess whether the echoes shown in the ringdown of linear radial perturbations survive also at the nonlinear level.

Our numerical setup is broadly the same as the one used in Ref.~\cite{Corelli:2022phw}, therefore here we will only highlight the main procedures.
The dimensional quantities will be expressed in terms of the mass scale $M_0$, which represents the mass of the scalarized BHs at the beginning of our simulations.

\subsection{Numerical setup}

The metric is expressed using Painlevé-Gullstrand-like (PG-like) coordinates, in the form:
\begin{equation}
ds^2 = -\hat\alpha^2 dt^2 + (R' dr + \hat\alpha \zeta \, dt)^2 + R^2 d\Omega^2,
    \label{eq:PG_ansatz}
\end{equation}
where $\hat\alpha(t, r)$ and $\zeta(t, r)$ are metric functions, while $R(r)$ is a radial ``fisheye'' transformation from the coordinate radius $r$ to the areal radius $R$. Its purpose is to allow to have a nonuniform resolution in $R$ starting from a uniform spatial discretization in $r$. This is particularly convenient for our purposes, since the BH solutions we are interested in have a small horizon areal radius, and require a small grid step to be resolved. By an appropriate choice of $R(r)$ we can decrease the resolution far from the BH, reducing substantially the computational cost of the simulations. In particular, the profile of $R(r)$ we considered is (see~\cite{Corelli:2022phw} for details)
\begin{align} 
        R &=  \eta_2 r + \frac{1 - \eta_1}{\Delta} \ln \biggl( \frac{1 + e^{-\Delta (r - r_1)}}{1 + e^{\Delta r_1}} \biggr) \notag  \\
          &+ \frac{\eta_2 - 1}{\Delta} \ln \biggl( \frac{1 + e^{-\Delta (r - r_2)}}{1 + e^{\Delta r_2}} \biggr),
        \label{eq:RTildeProfile}
\end{align}
where $r_{1,2}$ identify the locations where the resolution in $R$ changes, and $\Delta^{-1}$ their typical width, while $\eta_1$ and $\eta_2$ are the resolution gains in $r \lesssim r_1$ and $r \gtrsim r_2$, respectively. In other words, if $\Delta r$ is the grid step in the coordinate radius $r$, and $\Delta R$ the corresponding resolution in the areal radius, we will have that $\Delta R = \eta_1 \Delta r$ for $r \ll r_1$, and $\Delta R = \eta_2 \Delta r$ for $r \gg r_2$. This can be better appreciated from the derivative $R'(r)$, whose representative profile is is shown in Appendix~C of Ref.~\cite{Corelli:2022phw}.  In our simulations, the parameters in~\eqref{eq:RTildeProfile} have been set to
\begin{align}
    \eta_1 = 0.1, \quad & ~ \eta_2 = 10, \notag \\
    \quad r_1 = 10 \, M_0, \quad & ~ r_2 = 50 \, M_0\notag \\
    \Delta ~ =& ~~ 1 / M_0.\label{eq:RTParameters}
\end{align}

The electromagnetic tensor can be expressed using a 3+1 decomposition (see Ref.~\cite{Alcubierre:2009ij}). Since our setup is spherically symmetric the magnetic field vanishes, and $F^{\mu\nu}$ reads
\begin{equation}
    F^{\mu\nu} = n^\mu E^\nu - n^\nu E^\mu,
    \label{eq:FmunuDecomposition}
\end{equation}
where $n^\mu$ is the 4-velocity of the Eulerian observer, and $E^\mu = (0, E^r(t, r), 0, 0)$ is the electric field. As for the scalar field, we have $\phi = \phi(t, r)$; furthermore we introduce the auxiliary variables
\begin{equation}
    \Theta = \rder \phi, \qquad P = \frac{1}{\hat\alpha} \tder \phi - \frac{\zeta \Theta }{R'},
    \label{eq:ThetaPdefinitions}
\end{equation}
where $R' = \frac{d R}{d r}$. Plugging these definitions and the ansatz for the metric~\eqref{eq:PG_ansatz} into the field equations~\eqref{eq:scalar_eq} -~\eqref{eq:Einstein_eq}, we obtain the following system of partial differential equations
\begin{align}
    \tder \phi &= \hat\alpha \biggl( P + \frac{\zeta \Theta}{R'} \biggr) \label{eq:NonlinearPhiEvol} , \\
    \tder \Theta &= P \rder \hat\alpha + \hat\alpha \rder P + \frac{\zeta \Theta \rder \hat\alpha +  \hat\alpha \Theta \rder \zeta + \hat\alpha \zeta \rder \Theta}{R'} \notag \\
            &- \frac{\hat\alpha \zeta \Theta R''}{R'^2}  \label{eq:NonlinearThetaEvol}, \\
    \tder P &= \frac{\hat\alpha}{2 R \zeta R'^3} \Bigl\{ -\Theta^2 R^2 R' P + 2 \Theta \zeta \bigl(2 R'^2 - R R'' \bigr) \notag \\
            &+ R' \bigl[ R^2 P R'^2 \bigl(P^2 + F[\phi] (E^r)^2 R'^2 \bigr) \notag \\ 
            &+ R\zeta \Bigl((E^r)^2 F'[\phi] R'^4 + 2 \rder \Theta \Bigr) \notag \\
            &+ \zeta^2 R' (3 P R' + 2 R \rder P) \bigr] \Bigr\} \label{eq:NonlinearPEvol} , \\
    \tder \zeta &= \frac{\hat\alpha}{2} \biggl[ \frac{\zeta^2}{R} +R \biggl( P^2 + \frac{\Theta^2}{R'^2} + \frac{2 \Theta P}{\zeta R'} \notag \\
                &- F[\phi] (E^r)^2 R'^2 \biggr) + \frac{2 \zeta \rder \zeta}{R'} \biggr] , \label{eq:NonlinearZetaEvol} \\
    \tder E^r &= -\frac{\hat\alpha E^r F'[\phi]}{F[\phi]} \Bigl( P + \frac{\Theta \zeta}{R'}\Bigr) \label{eq:NonlinearErEvol},
\end{align}
where $F'[\phi] = \frac{\delta F[\phi]}{\delta \phi}$, and the following constraints
\begin{align}
    \rder \zeta &= - \frac{\zeta R'}{2 R} + R P \Theta + \frac{R}{2 \zeta R'} \Bigl[ P^2 R'^2 + \Theta^2 \notag \\
                &+ F[\phi] (E^r)^2 R'^4 \Bigr], \label{eq:NonlinearZetaConstraint} \\
    \rder \hat\alpha &= - \frac{\hat\alpha R P \Theta}{\zeta} \label{eq:NonlinearAlphaConstraint} , \\
    \rder E^r &= -E^r \biggl( \frac{2 R'}{R} + \frac{R''}{R'} + \frac{F'[\phi] \Theta}{F[\phi]} \biggr). \label{eq:NonlinearErConstraint}
\end{align}

To perform the evolution we adopt the method of lines, computing spatial derivatives with the finite difference formula at the fourth order of accuracy, and using the fourth-order accurate Runge-Kutta method for time integration. At each iteration, we evolve $\phi$, $\Theta$, $P$, $\zeta$ and $E^r$ with Eqs.~\eqref{eq:NonlinearPhiEvol}--\eqref{eq:NonlinearErEvol}, and then we integrate Eq.~\eqref{eq:NonlinearAlphaConstraint} to obtain $\hat\alpha$. In particular, since this latter equation can be written as
\begin{equation}
    \rder \ln \hat\alpha = - \frac{R P \Theta}{\zeta},
    \label{eq:NonlinearLogAlphaConstraint}
\end{equation}
we first use a combination of the Simpson's rules to integrate for $\ln \hat\alpha$, and then we exponentiate (c.f. Ref.~\cite{Corelli:2022phw}). To prevent high-frequency instabilities we include a fifth order Kreiss-Oliger dissipation term in the right-hand side of the evolution equations~\eqref{eq:NonlinearPhiEvol} -~\eqref{eq:NonlinearErEvol}. 

Inside the BH we perform excision, removing from the domain of integration the region in the interior of a given radius $r_e$, which is set to the be a quarter of the horizon radius at the beginning of the simulation and is never updated. In the first 3 grid points outside the excision radius we compute the numerical derivatives using the one-sided formula, and we perform the evolution as in the rest of the domain\footnote{In our implementation of the integration algorithm, it is the fourth grid point (the first in which the centered formula for derivation is used) that is set at $R_h/(4 \eta_1)$. Here $R_h$ is the areal radius of the horizon of the static BH we start with, and since we are at radii smaller than $r_1$ in Eq.~\eqref{eq:RTParameters}, $R_h/\eta_1$ represents an estimate of the coordinate value of the horizon radius. The actual location of the excision radius is then taken to be 3 grid steps before $R_h/(4 \eta_1)$.}. At the outer boundary instead we add 3 ghost zones where the fields are kept constant throughout the simulation. 

To track the position of the apparent horizon, we search for the radius where $\zeta = 1$, which is the location at which the expansion of future-directed radial null geodesics vanishes.

\subsection{Initial data}

The initial condition of our simulations is composed by a scalarized BH with an infalling perturbation of the scalar field. 

To build such initial setup, we start constructing a static scalarized BH solution in PG-like coordinates. For this purpose we remove the time-dependence from all the fields, and we consider an ansatz for the metric that does not include the radial transformation:
\begin{equation}
    ds^2 = -\hat\alpha(r)^2 dt^2 + (dR + \hat\alpha(r) \zeta(r) \, dt)^2 + R^2 d\Omega^2.
    \label{eq:PG_ansatz_static}    
\end{equation}
Substituting it in the field equations, we obtain 
\begin{align}
    \Rder \zeta &= \frac{1}{2 R \zeta} \Bigl\{ R^2 \Bigl[F[\phi] \bigl(E^R \bigr)^2 + \bigl(\Rder \phi \bigr)^2 \Bigr] \notag \\
                &- \zeta^2 \Bigl[1 + R^2 \bigl(\Rder \phi \bigr)^2 \Bigr]\Bigr\} \label{eq:StaticPGZeta} \\
    \Rder^2 \phi &= -\frac{(\zeta^2 - 2) \Rder \phi}{R(\zeta^2 - 1)} + \frac{\bigl(E^R\bigr)^2 \bigl(F'[\phi] - 2 R F[\phi]\Rder \phi \bigr)}{2(\zeta^2 - 1)}. \label{eq:StaticPGphi}
\end{align}
Note that here we do not use the auxiliary variables $\Theta$ and $P$, but we work directly with the scalar field. Furthermore, since $\hat\alpha$ does not appear in the equations for $\phi$ and $\zeta$, we do not integrate for it. For the electric field the solution has closed form: $E^R = \frac{Q}{R^2 F[\phi]}$, where $Q$ is the electric charge.

At the horizon $\zeta = 1$, and the denominator of Eq.\eqref{eq:StaticPGphi} vanishes. The requirement of regularity at the horizon thus yields the relation
\begin{equation}
    \Rder \phi_h = \frac{R_h \bigl(E^R_h\bigr)^2 F'[\phi_h]}{2 \Bigl(-1 + R_h^2 F[\phi_h] \bigl(E^R_h\bigr)^2\Bigr)},
    \label{eq:StaticPGHorizonCondition}
\end{equation}
where the subscript $h$ indicates that the quantities are evaluated at the horizon. Substituting this condition in Eq.~\eqref{eq:StaticPGphi} we obtain
\begin{equation}
    \Rder^2 \phi_h = \frac{\bigl(E^R_h\bigr)^2 F'[\phi_h]}{2 \Bigl(1 - R_h^2 F[\phi_h] \bigl(E^R_h\bigr)^2 \Bigr)}.
    \label{eq:StaticPGPhidderHorizon}
\end{equation}

At infinity instead we have the expansions  
\begin{align}
    \phi = \frac{Q_s}{R} + \OO{\frac{1}{R^2}}, \quad \zeta = \sqrt{\frac{2 M}{R}} + \OO{\frac{1}{R^{3/2}}},
    \label{eq:StaticPGAsymptotic}
\end{align}
where $Q_s$ is the scalar charge, and $M$ is the asymptotic value of the Misner-Sharp mass function.

Starting from a fixed value of the horizon areal radius and the electric charge, we construct the solution via a shooting procedure that searches for the value of $\phi_h$ that leads to asymptotic behavior for the scalar field in Eq.~\eqref{eq:StaticPGAsymptotic}.  At the horizon we use Eq.~\eqref{eq:StaticPGPhidderHorizon} to obtain the right-hand side of the equation for $\Rder^2 \phi$, and we perform the integration for a grid step that is half the grid step used in the rest of the numerical grid. After the shooting we rescale the electric charge and all the numerical grid by the BH mass, then we perform the shooting procedure again; in this way the solution will have unitary mass. Finally, we construct the full solution by integrating both outward and inward, and computing the conjugate momentum of the scalar field as $P = -\zeta \Rder \phi$. On the horizon we use a halved grid step, in such a way that at the end the horizon is placed between two grid points. In all the procedure numerical integration is performed with the fourth-order accurate Runge-Kutta method, and the resolution of the final solution is considerably higher than the one used in the time evolution. 

Once the BH solution has been obtained, we add the perturbation on top of it. In particular, the initial profile of the scalar field in the simulation is set to
\begin{align}
    \phi(r, t = 0) &= \phi_0(R(r)) + \delta \phi(R(r)), \notag \\
    \Theta(r, t = 0) &= R'(r) \Bigl( \Theta_0(R(r)) + \Rder \delta \phi (R(r)) \Bigr), \notag \\
    P(r, t = 0) &= P_0(R(r)) + \frac{\delta \phi(R(r))}{R(r)} + \Rder \delta \phi(R(r)),
    \label{eq:InitialScalarField}
\end{align}
where $\phi_0$, $\Theta_0$ and $P_0$ denote the profile of the scalarized solution. $\delta \phi$ instead is a Gaussian wave packet:
\begin{equation}
    \delta \phi(R) = \frac{A}{R} e^{-\frac{(R - R_0)^2}{2 \sigma^2}} \cos[k (R - R_0)],
    \label{eq:ScalarWavePacket}
\end{equation}
where $A$ is its amplitude, $R_0$ its location, $\sigma$ its typical width, and $k$ its wave number. The profile of the perturbation in $P$ is the one of an ingoing wave in flat spacetime, chosen in such a way to impose that a large component of the initial wave packet is directed towards the BH.

Since the scalarized BH solution is computed with a different resolution than the one used in numerical integration, the values of $\phi_0$, $\Theta_0$ and $P_0$ on the numerical grid are obtained through a fourth-order interpolation with Lagrange polynomials. To ensure that the constraints are satisfied on initial data we set the profiles of the electric field and metric functions from a numerical integration of the constraints~\eqref{eq:NonlinearZetaConstraint} -~\eqref{eq:NonlinearErConstraint} that starts from the first grid point outside the horizon and proceeds both outward and inward. As starting values for $E^r$ and $\zeta$ we use those obtained from the scalarized BH, reflecting the idea that the horizon structure is not affected by the presence of the wave packet, if the latter is distant enough. As for $\hat\alpha$, we set its starting value to $1$, and at the end of the procedure we rescale the whole profile in such a way that its value at the outermost grid point (before the ghost zones) is unity. This operation is allowed by the residual gauge freedom in PG-like coordinates, and can be interpreted as the requirement that coordinate time coincide with proper time at spatial infinity.
The entire procedure to compute the initial data is similar to the one implemented in previous work (see, e.g.,~\cite{Corelli:2022phw}).

\subsection{Numerical results} \label{sec:NonlinearResults}

We now present the results of our numerical simulations. As a first step we reproduced the echo patterns obtained in the linear regime by considering a small value for the amplitude of the scalar perturbation, $A = 0.001 \, M_0$. The other parameters in Eq.~\eqref{eq:ScalarWavePacket} were set to $r_0 = 10 \, M_0$, $\sigma = M_0$ and $k = 0$. We considered the two cases shown in Fig.~\ref{fig:TD_linear}: $\alpha = 0.5$ with $q = 1.0073$, and $\alpha = 0.6$ with $q = 1.0187$. In all the simulations the grid step was chosen to be $\Delta r = \frac{r_{h, 0} - r_{e, 4}}{140}$, where $r_{h, 0}$ is the coordinate horizon radius at the beginning of the simulation, and $r_{e, 4}$ is the coordinate radius of the first grid point where the centered formula for derivation is used. The time step was set to $\Delta t = 0.2 ~ \eta_1 ~ \Delta r$, where $\eta_1$ is the parameter entering in the transformation law~\eqref{eq:RTildeProfile}, which also represents the inverse of the coordinate propagation speed of signals in the internal region $r \lesssim r_1$. The outer boundary was placed at $r_\infty = 210 \, M_0$, corresponding to an areal radius $R_\infty = 1641 \, M_0$, while the final time of integration was set to $T = 1500 \, M_0$.

\begin{figure*}[th]
    \centering
    \includegraphics[width=0.49\linewidth]{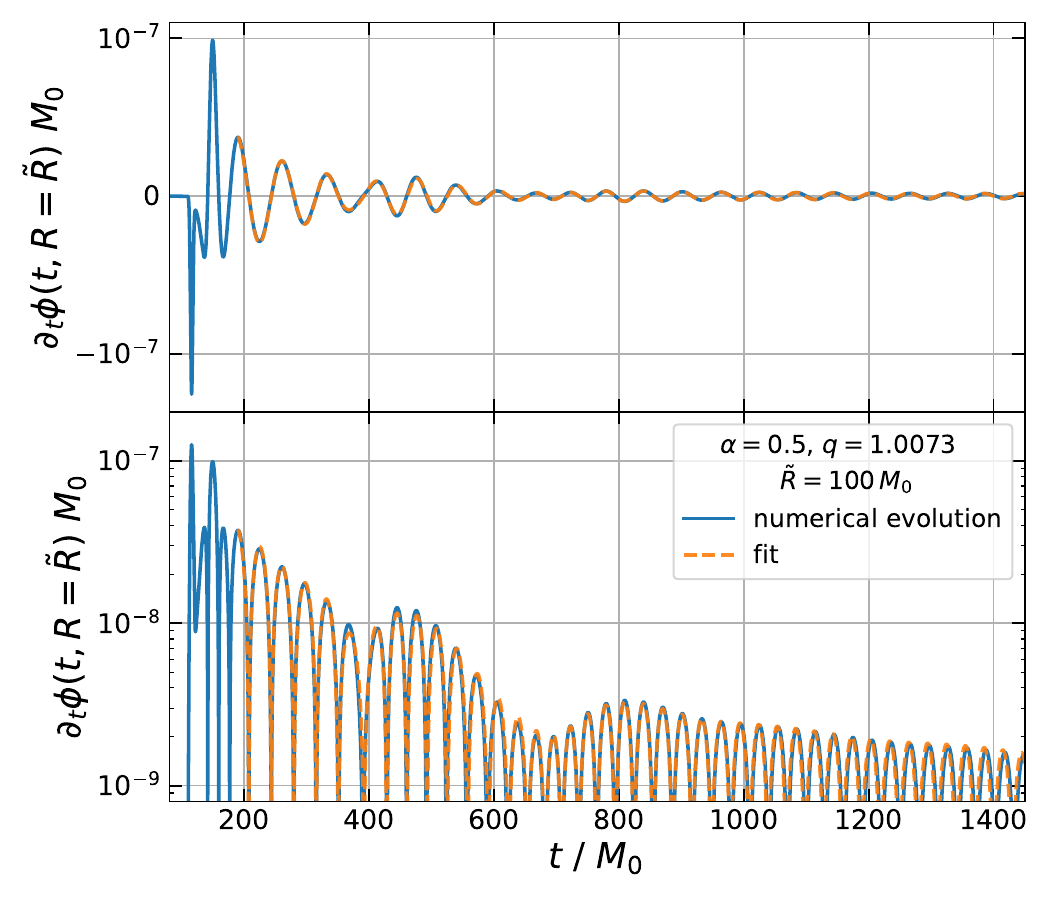} 
    \includegraphics[width=0.49\linewidth]{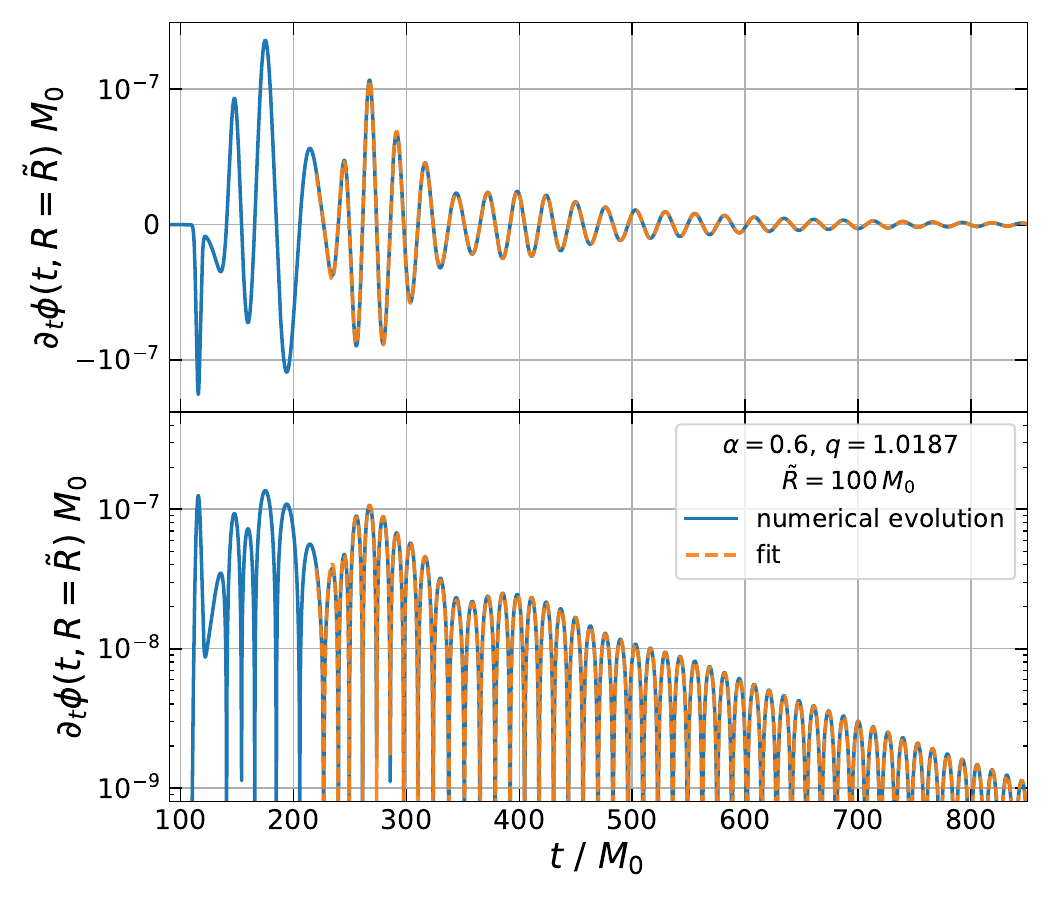}
    \caption{Asymptotic behavior of the time derivative of the scalar field extracted at $\tilde R = 100 \, M_0$ from the simulations of the collapse of wave packets of $\phi$ onto scalarized BHs. The left panel shows the case with $\alpha = 0.5$ and $q = 1.0073$, while the right panel the case with $\alpha = 0.6$ and $q = 1.0187$. In both simulations the amplitude of the scalar perturbation was set to $A = 0.001 \, M_0$, so that the dynamics is in the linear regime; nonetheless, numerical evolution is performed at the fully-nonlinear level. Blue solid lines represent the behavior of $\tder \phi(t, \tilde R)$, while orange dashed lines represent the fit with four damped sinusoids (see Eq.~\eqref{superposition}) whose complex frequencies are the ones reported in Fig.~\ref{fig:TD_linear}. The fits start at $t = 190 \, M_0$ for the case with $\alpha = 0.5$, and at $t = 220 \, M_0$ for the case with $\alpha = 0.6$. In both plots the top and bottom panels show the same profiles in linear and semi-log scales, respectively.}
    \label{fig:NonlinearSimulationsLinearRegime}
\end{figure*}

Throughout the simulations we extracted the scalar field at $\tilde r = 55.9 \, M_0$, corresponding to $\tilde R = 100 \, M_0$, every $2048$ time steps. Then we computed its time derivative numerically with a fourth-order accurate finite differences formula. This was done in order to remove the constant asymptotic value of the scalar field and isolate the behavior of the perturbation as the system is approaching the final scalarized configuration. We then fitted the time behavior of $\tder \phi(t, \tilde R)$ with a superposition of damped sinusoids (see Eq.~\eqref{superposition}) whose complex frequencies are those computed at the linear level in the frequency domain, and reported in Fig.~\ref{fig:TD_linear}. The results for the cases $\alpha = 0.5$ and $\alpha = 0.6$ are shown, respectively, in the left and right plots of Fig.~\ref{fig:NonlinearSimulationsLinearRegime}, with the behaviors of $\tder \phi(t, \tilde R)$ being represented by blue solid curves, and the fits by orange dashed lines. In each of the two plots we display the results in linear scale in the top panel, and in semi-log scale in the bottom one. The fits start at $t = 190 \, M_0$ and $t = 220 \, M_0$ for the simulation with $\alpha = 0.5$ and $\alpha = 0.6$, respectively, and they reproduce well the asymptotic behavior of $\tder \phi(t, \tilde R)$. Furthermore we can clearly observe the same echo patterns found with a linear evolution code. 

Let us now discuss a case in which the dynamics is in the nonlinear regime. Note that when the wave packet is absorbed by the BH, it carries a positive contribution to the mass, leading to a decrease of the charge-to-mass ratio. Therefore, in order to obtain a final configuration that falls in the region of the parameter space where echoes are expected to appear, we need to start our simulation with a BH that is even closer to extremality, and has a considerably small horizon radius. As an example, we considered the case $\alpha = 0.6$, and constructed an initial configuration in which the BH has charge-to-mass ratio $q = 1.0194$ and horizon areal radius $R_h = 0.0199 \, M_0$; the amplitude of the scalar wave packet was set to $A = 0.03 \, M_0$, and the other parameters to $r_0 = 10 \, M_0$, $\sigma = M_0$, $k = 0$. As in the linear case, we set the grid step to $\Delta r = \frac{r_{h, 0} - r_{e, 4}}{140}$, and the time step to $\Delta t = 0.2 ~ \eta_1 ~ \Delta r$. The outer boundary was placed at $r_\infty = 210 \, M_0$, and the final time of integration was set to $T = 1500 \, M_0$. 

To determine the final BH configuration, and the QNM frequencies associated to it, we proceeded in the following way: we extracted the horizon areal radius at the end of the simulation, $R_h$, using a fourth-order polynomial interpolation to determine the location where the expansion vanishes; then we computed the horizon charge $Q_h$ with a fourth-order polynomial interpolation of $Q(T, r) = E^r (T, r) F[\phi(T, r)] R^2 R'$ at the horizon; we used a shooting procedure to construct the static BH solution with radius $R_h$ and charge $Q_h$, adopting $\phi(T, r_h)$ (also computed via interpolation) as initial value, and we used it as an approximation of the final configuration, from which we computed mass and charge-to-mass ratio. In our simulation we obtained a final mass $M = 1.001 \, M_0$, corresponding to a $q = 1.0182$. Even if the increase in mass is only $\approx 0.1\%$, the system is well in the nonlinear regime, as the areal radius has grown by a factor $\approx 5.7$, reaching a final value $R_h = 0.114 \, M_0$ (and correspondingly the horizon area has grown by a factor $\approx 32$). 

\begin{figure}
    \centering
    \includegraphics[width=\columnwidth]{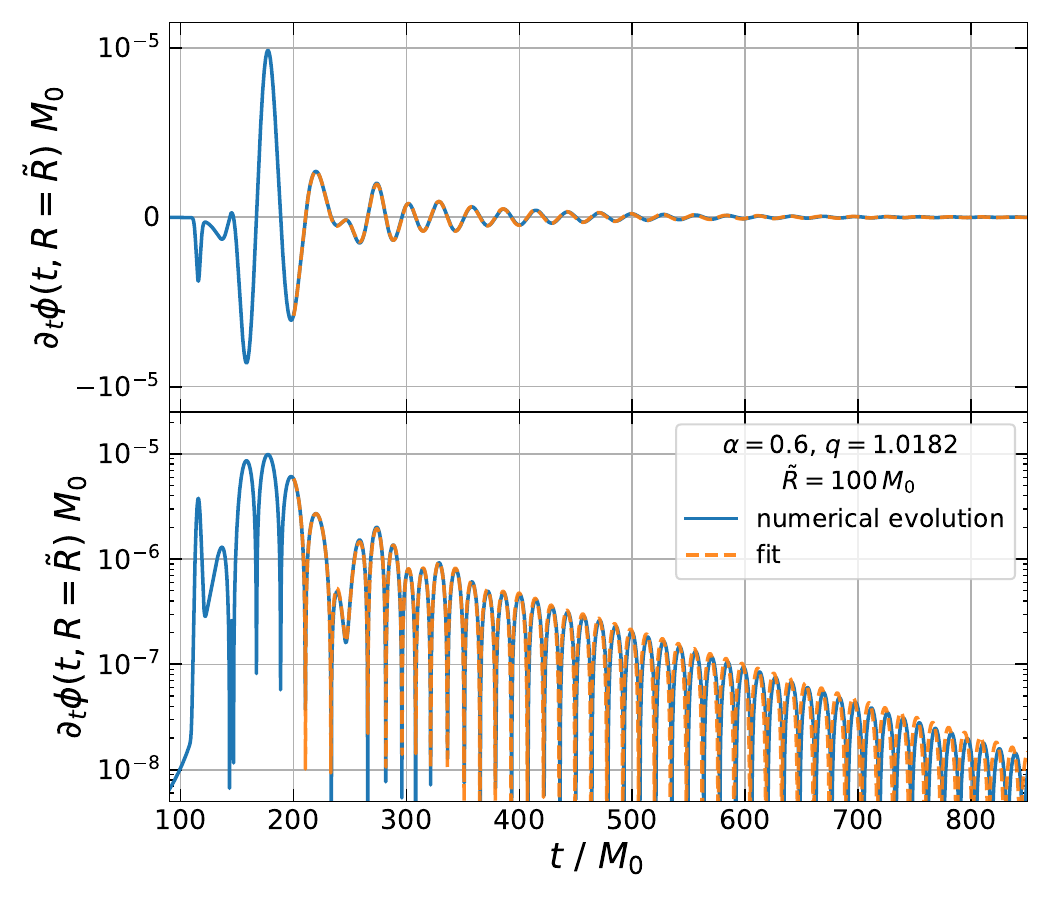}
    \caption{Asymptotic behavior of the time derivative of the scalar field at $\tilde R = 100 \, M_0$ for the simulation in which the dynamics is in the nonlinear regime. Conventions are consistent with Fig.~\ref{fig:NonlinearSimulationsLinearRegime}, and the fit is performed for $t > 200 \, M_0$. Even if during the simulation the horizon radius has increased by a factor $\approx 5.7$ (horizon area increased by a factor $\approx 32$), echoes appear and the asymptotic behavior is well-approximated by a superposition of the first 4 QNMs of the final BH configuration.}
    \label{fig:SimulationNonlinearAlpha0.6}
\end{figure}

Once the final configuration has been obtained, we extracted the complex frequencies of the first 4 QNMs associated to it by linear interpolation from the set computed in Sec.~\ref{sec:QNM_linear_radial}. We then repeated the analysis we performed for the linear case, and we plotted the results in Fig.~\ref{fig:SimulationNonlinearAlpha0.6}. As we can see, echoes are not destroyed and the superposition of QNMs of the final BH well approximates the asymptotic behavior of $\tder \phi(t, \tilde R)$. This also shows consistency between the frequency domain computation in the linear regime and the time domain analysis at the nonlinear level.

\section{Conclusion and extensions}

In this work, we investigated the behavior of BHs in EMS theory, focusing on both linear and nonlinear dynamics. This theory, characterized by a nonminimal coupling between the electromagnetic and scalar fields, admits spherically symmetric BH solutions with a stable photon sphere. These configurations yield effective potentials with multiple extrema, leading to a rich QNM spectrum and the emergence of long-lived modes trapped within potential cavities.

A key result of our study is the demonstration of echo signals in the linear response of these BHs, similar to what has been previously observed for horizonless compact objects. However, our work goes beyond linear perturbation theory, showing that echoes persist even in fully nonlinear regimes. Through 1+1 numerical simulations, we provided the first example of such a phenomenon in a consistent theoretical framework that includes nonlinear effects. This result underscores the robustness of echoes as a feature of this model, and the potential of EMS theory as a 
playground for exploring large deviations from standard BH dynamics. This also confirms that GW BH spectroscopy can serve as a powerful tool for probing exotic phenomena and testing modifications of GR.

Overall, the echo pattern found for both linear perturbations and nonlinear dynamical evolution is less evident than the characteristic echoes of exotic compact objects~\cite{Cardoso:2016rao,Cardoso:2016oxy,Cardoso:2017cqb}. Indeed, in our case echoes are more evident in the logarithmic scale shown in the bottom panels of Figs.~\ref{fig:NonlinearSimulationsLinearRegime} and \ref{fig:SimulationNonlinearAlpha0.6}. This difference arises because the confinement mechanism of the scalarized BHs analyzed in this manuscript is less efficient than in the case of ultracompact objects, where the inner boundary is fully or highly reflective. For the scalarized BHs we studied, instead of having a reflective surface at the inner boundary there is an horizon that, as a dissipation channel, contributes to weaken subsequent echoes. Furthermore, the articulated shape of the effective potential contributes to making the echo pattern more complex.

Our findings pave the way for further investigations into the post-merger GW signal from BH collisions in EMS theory. It remains to be seen whether the echoes observed in this work at the nonlinear level survive also in the nonlinear conditions of a coalescence. This would also provide an arena to check and quantify certain arguments against echoes at the nonlinear level, for example due to energy loss~\cite{Vellucci:2022hpl,Dailey:2023mvn} or to the
expansion of an apparent horizon~\cite{Guo:2022umn} that might prevent potential reflections or make them less efficient (see \cite{Dailey:2023mvn} for a critical discussion).

Finally, this framework might be used also to study the photon-sphere instability at the nonlinear level. Since our stationary solutions are BHs, the hypotheses of the theorem presented in~\cite{Cunha:2017qtt} do not apply, and indeed we did not find any signatures of an instability in our nonlinear radial simulations. It remains to be seen whether this is a generic feature and whether it is valid also for nonradial perturbations.
We will come back to these problems in a follow-up work~\cite{echoesnonlinPreliminary}.

\begin{acknowledgements}
This work is partially supported by the MUR PRIN Grant 2020KR4KN2 ``String Theory as a bridge between Gauge Theories and Quantum Gravity'', by the FARE programme (GW-NEXT, CUP:~B84I20000100001), and by the INFN TEONGRAV initiative.
Some numerical computations have been performed at the Vera cluster supported by the Italian Ministry for Research and by Sapienza University of Rome.
\end{acknowledgements}

\appendix 

\section{Derivation of the perturbed field equations}\label{app:App_der_pert}
In this Appendix we outline the main steps to derive the perturbed field equations~\eqref{eq:axial_U}-\eqref{eq:axial_H} and~\eqref{eq:polar_f12_gen}-\eqref{eq:polar_z_gen}. The linearized Einstein, Maxwell and scalar equations can be divided into three distinct groups (“scalar'', “vector'' and “tensor'') as described in~\cite{PhysRevD.46.4289} (see also~\cite{Brito:2018hjh}). By denoting the linearized Einstein, Maxwell and scalar equations respectively as $\delta \mathcal{E}_{\mu\nu} = 0$, $\delta J_\mu = 0$ and $\delta S = 0$, the “scalar'' group is given by the equations
\begin{align}
    \delta \mathcal{E}_{(I)} =& \sum_{l,m} A_{lm}^{(I)}(r) \, Y_l^m \;, \hspace{1cm} (I=0, \dots, 3) \\
    \delta J_{(I)} =& \sum_{l,m} A_{lm}^{(I)}(r) \, Y_l^m \;, \hspace{1cm} (I=4,5) \\
    \delta S =& \sum_{l,m} A_{lm}^{(6)}(r) \, Y_l^m \;,
\end{align}
where $I=0,1,2$ stand for the $(tt)$, $(tr)$ and $(rr)$ components of the Einstein equations, $I=4$ represents the following combination
\begin{equation}
    \delta \mathcal{E}_{\theta \theta} + \frac{\delta \mathcal{E}_{\varphi \varphi}}{\sin^2 \theta} = 0 \;,
\end{equation}
and $I=4,5$ indicate the $(t)$ and $(r)$ components of the Maxwell equations.
The “vector'' group is represented by the following Einstein equations 
\begin{align}
    \delta \mathcal{E}_{(L \theta)} &= \alpha_l^{(L)}(r) \, \partial_\theta Y_l^m - \beta_l^{(L)}(r) \, \frac{\partial_\varphi Y_l^m}{\sin \theta} = 0 \;, \\
    \frac{\delta \mathcal{E}_{(L \varphi)}}{\sin \theta} &= \beta_l^{(L)}(r) \, \partial_\theta Y_l^m + \alpha_l^{(L)}(r) \, \frac{\partial_\varphi}{\sin\theta} = 0 \;,
\end{align}
where $L=0,1$ stand for $(t)$ and $(r)$, and by the remaining Maxwell equations
\begin{align}
    \delta J_\theta &= \alpha_l^{(2)}(r)\, \partial_\theta Y_l^m - \beta_l^{(2)}(r)\, \frac{\partial_\varphi Y_l^m}{\sin \theta} = 0 \;, \\
    \frac{\delta J_\varphi}{\sin\theta} &= \beta_l^{(2)}(r)\, \partial_\theta Y_l^m + \alpha_l^{(2)}(r)\, \frac{\partial_\varphi Y_l^m}{\sin \theta} = 0 \;.
\end{align}
The “tensor'' group consists of the remaining Einstein equations
\begin{align}
    \frac{\delta \mathcal{E}_{\theta \varphi}}{\sin\theta} &= s_l(r) \frac{X^l}{\sin\theta} + t_l(r) W^l = 0 \;, \\
    \delta \mathcal{E}_{\theta \theta} + \frac{\delta \mathcal{E}_{\varphi \varphi}}{\sin^2 \theta} &= -t_l(r) \frac{X^l}{\sin\theta} + s_l(r) W^l = 0 \;, 
\end{align}
where $X^l$ and $W^l$ are so defined
\begin{align}
    X^l &= 2 \partial_\varphi (\partial_\theta Y_l^m - \cot\theta Y_l^m) \;, \\
    W^l &= \partial^2_\theta Y_l^m -\cot\theta \partial_\theta Y_l^m - \frac{\partial^2_\varphi Y_l^m}{\sin^2\theta}.
\end{align}
The crucial point is that the coefficients $A_l^{(I)}$, $\alpha_l^{(L)}$, $\beta_l^{(L)}$, $s_l$ and $t_l$ are linear functions of the perturbations and purely radial. By using the orthogonality of the spherical harmonics it is possible to separate the radial from the angular part, to get the following equations that naturally divide into the axial sector
\begin{align}
    \begin{cases}
        \beta_l^{(L)} = 0 \hspace{0.5cm} L=0,1,2 \\
        t_l = 0 
    \end{cases}
\end{align}
with perturbation fields $u_4(r)$, $h_0(r)$ and $h_1(r)$, and polar sector 
\begin{align}
    \begin{cases}\label{eq:polar_der}
        A_l^{(L)} = 0 \hspace{0.5cm} I=0,\dots,6 \\
        \alpha_l^{(L)} = 0 \hspace{0.5cm} L=0,1,2 \\
        s_l = 0 \;,
    \end{cases}
\end{align}
with perturbation fields $H_0(r)$, $H_1(r)$, $H_2(r)$, $K(r)$, $f_{01}(r)$, $f_{02}(r)$, $f_{12}(r)$ and $z(r)$. 

In order to obtain the final form of the axial equations~\eqref{eq:axial_U}-\eqref{eq:axial_H}, we first consider the definitions 
\begin{equation}
    h_1(r) = \frac{Q_1(r)}{N(r)}\;, \hspace{1cm} u_4(r) = \frac{i U_4(r)}{\omega} \;.
\end{equation}
From the equation $t_l=0$ we obtain the following relation
\begin{equation}
    h_0(r) = \frac{i N(r)}{\omega e^{2 \delta(r)}} [Q_1'(r) - Q_1(r)\delta'(r)]
\end{equation}
that removes $h_0(r)$ from the system. It is easy to prove that equation $\beta_l^{(0)} = 0$ is redundant because it can be derived from $\beta_l^{(1)} = 0$ and $\beta_l^{(2)} = 0$. We are left with a system of two differential equations $\beta_l^{(1)}=0$ and $\beta_l^{(2)}=0$, where for convenience we replace the latter with the following linear combination
\begin{equation}
    \beta_l^{(2)} + \frac{2 N(r) V'(r) \, e^{\alpha \phi^2(r)}}{\omega} \beta_l^{(1)} = 0 \;.
\end{equation}
Finally, using a further redefinition given by 
\begin{equation}\label{eq:U_H_defs}
    U_4(r) = r \, e^{-\frac12 \alpha \phi^2(r)} \, U(r)\;, \hspace{0.5cm} Q_1(r) = r \, e^{\delta(r)} H(r) \;,
\end{equation}
we obtain the desired form of the axial perturbed equations. 

The polar equations~\eqref{eq:polar_der}, after eliminating $f_{01}(r)$ with Eq.~\eqref{eq:bianchi} and deriving $H_2(r) = H_0(r)$ from $s_l=0$, directly give the Eqs.~\eqref{eq:polar_f12_gen}-\eqref{eq:polar_z_gen}.

\section{Convergence tests on nonlinear 1+1 evolution code}

To test the accuracy of our fully-nonlinear 1+1 code, we checked how the violations of the constraints scale with resolution. In particular we considered the violations of~\eqref{eq:NonlinearZetaConstraint} and~\eqref{eq:NonlinearErConstraint}, which are, respectively, 
\begin{align}
    CV_\zeta &=  \rder \zeta  + \frac{\zeta R'}{2 R} - R P \Theta - \frac{R}{2 \zeta R'} \Bigl[ P^2 R'^2 + \Theta^2 \notag \\
             &+ F[\phi] (E^r)^2 R'^4 \Bigr], \label{eq:CVzeta}
\end{align}
and
\begin{equation}
    CV_{E^r} = \rder E^r + E^r \biggl( \frac{2 R'}{R} + \frac{R''}{R'} + \frac{F'[\phi] \Theta}{F[\phi]} \biggr). \label{eq:CVGauss}
\end{equation}

Since in our code we use fourth-order accurate procedures, both for computing spatial derivatives and for time integration, we expect that constraint violations $CV^{(\Delta r_1)}$ and $CV^{(\Delta r_2)}$, corresponding respectively to simulations with grid step $\Delta r_1$ and $\Delta r_2$, are related by
\begin{equation}
    CV^{(\Delta r_1)} = \left(\frac{\Delta r_1}{\Delta r_2} \right)^4 CV^{(\Delta r_2)}
    \label{eq:CVScalingExpected}
\end{equation}

\begin{figure}
    \centering
    \includegraphics[width=\columnwidth]{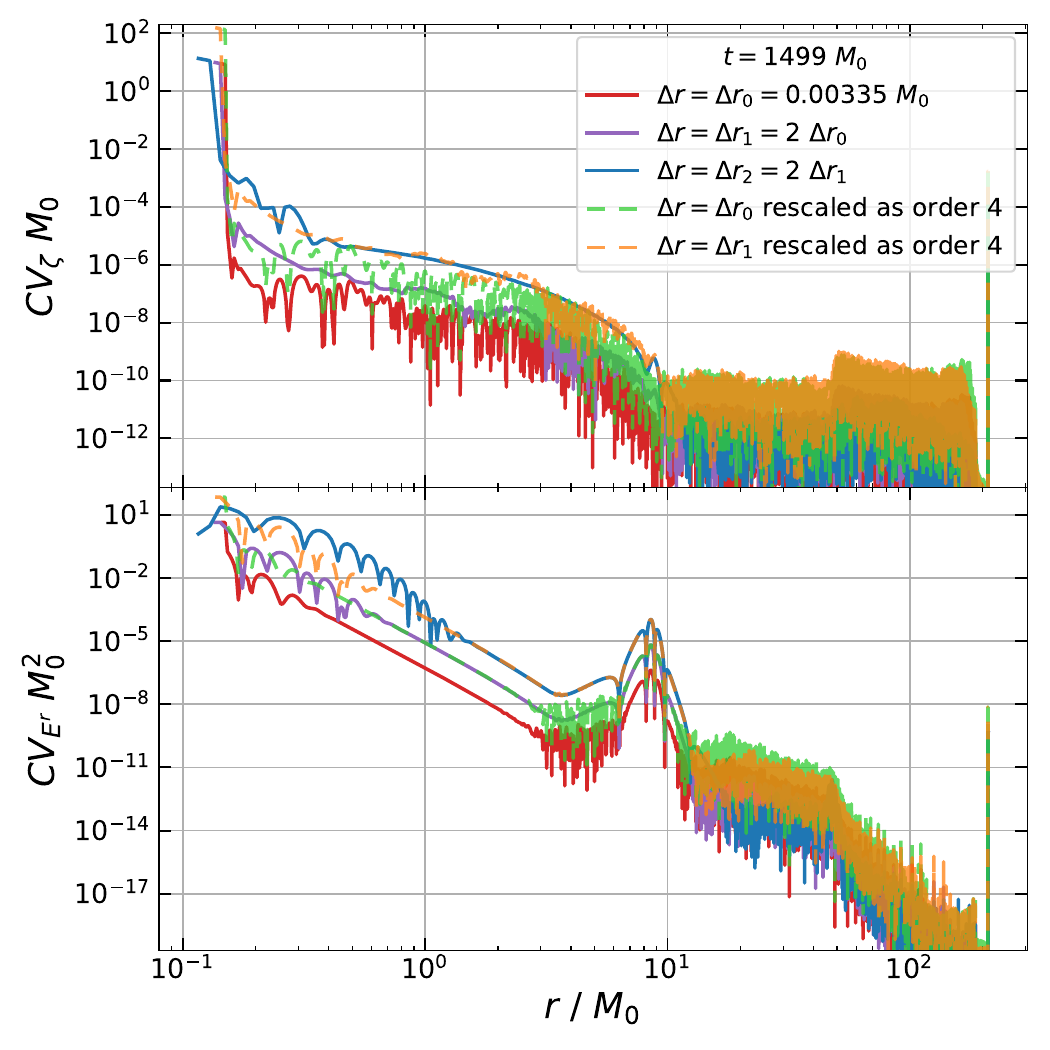}
    \caption{Scaling of the constraint violations for the simulation with $\alpha = 0.5$, in the case in which the dynamics is in the linear regime. Data are extracted at $t = 1499 \, M_0$, close to the end of the simulation. In the upper panel we show the scaling of~\eqref{eq:CVzeta}, and in the lower panel the scaling of~\eqref{eq:CVGauss}. Red, purple and blue solid lines denote the constraint violation extracted from the simulation with grid step $\Delta r_0$, $\Delta r_1$ and $\Delta r_2$, respectively, while the green and orange lines are the constraint violations corresponding to the simulations with $\Delta r_0$ and $\Delta r_1$ rescaled according to Eq.~\eqref{eq:CVScalingExpected}. We can see that the scaling behavior is generally in agreement with expectations, with the inner region entering in the convergence regime only at higher resolutions.}
    \label{fig:Alpha0.5_linear_convergence}
\end{figure}

\begin{figure}
    \centering
    \includegraphics[width=\columnwidth]{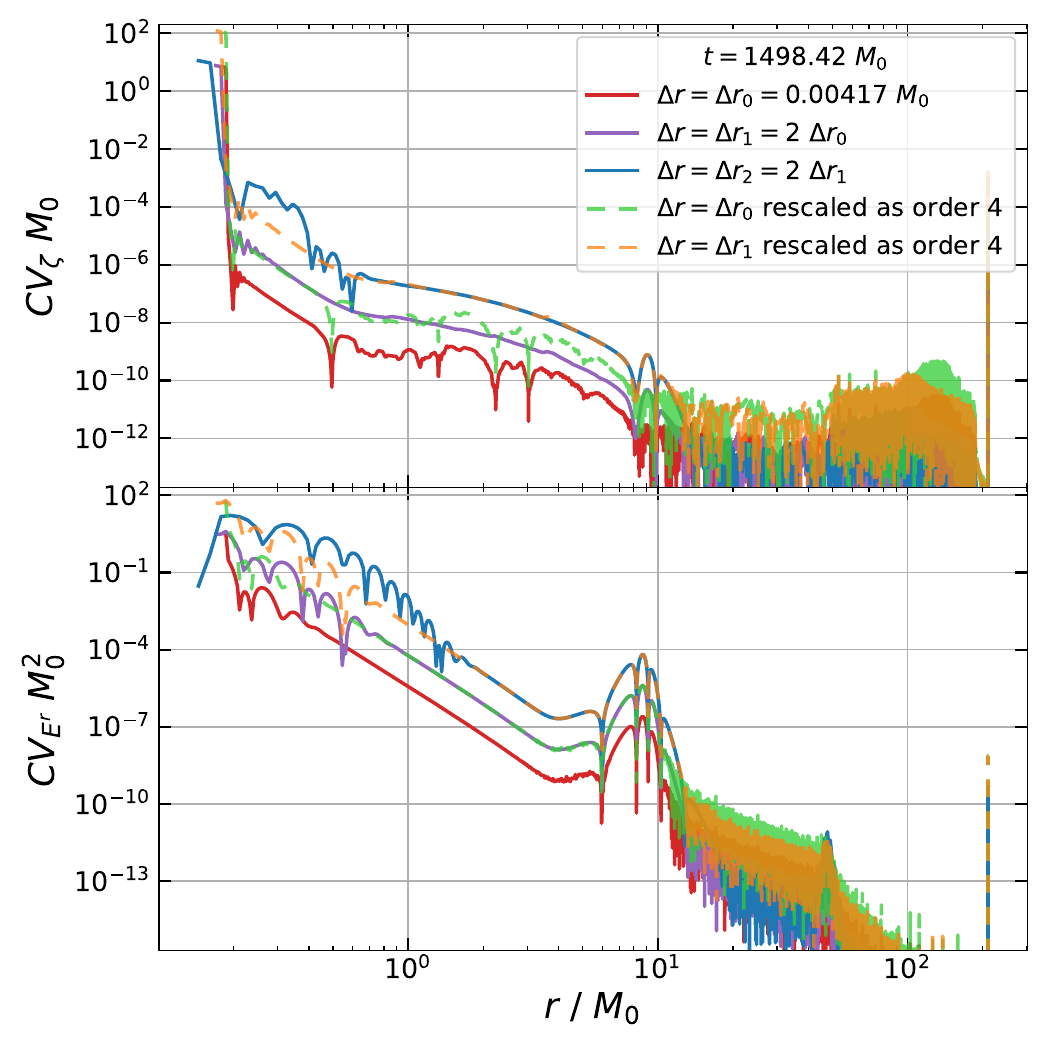}
    \caption{Convergence of the constraint violations for the simulation with $\alpha = 0.6$ in which the dynamics is in the linear regime. Conventions are consistent with Fig.~\ref{fig:Alpha0.5_linear_convergence}, and we can see that also in this case the expected scaling behavior~\eqref{eq:CVScalingExpected} is generally satisfied. The constraint violations are generally larger in the BH region, and convergence is achieved at the higher resolutions.}
    \label{fig:Alpha0.6_linear_convergence}
\end{figure}

To check this scaling behavior we repeated each of the three simulations discussed in Sec.~\ref{sec:NonlinearResults} with a decreased resolution. Namely, if we denote with $\Delta r_0$ the grid step used in Sec.~\ref{sec:NonlinearResults}, we repeated the simulations twice, with $\Delta r_1 = 2 \Delta r_0$ and $\Delta r_2 = 2 \Delta r_1$ as grid steps, and increasing the time step $\Delta t$ by the same factors. We then computed the constraint violations~\eqref{eq:CVzeta} and~\eqref{eq:CVGauss} at a time close to the end of the simulations. The results are shown in Figs.~\ref{fig:Alpha0.5_linear_convergence} and~\ref{fig:Alpha0.6_linear_convergence} for the simulations in the linear regime with $\alpha = 0.5$ and $\alpha = 0.6$, respectively. In both plots the upper panel shows the scaling of $CV_{\zeta}$, while the lower panel the scaling of $CV_{E^r}$. Solid lines represent the constraint violations directly computed from the simulations with grid steps $\Delta r_0$ (red), $\Delta r_1$ (purple) and $\Delta r_2$ (blue), while dashed lines are the constraint violation from the simulations with $\Delta r = \Delta r_0$ (green) and $\Delta r = \Delta r_1$ (orange) rescaled according to Eq.~\eqref{eq:CVScalingExpected}. As we can see the value of the constraint violations is higher inside the horizon, and lower in the outer region. The convergence behavior is generally good, with the inner region entering in the convergence regime only at high resolutions. On the other hand in the region around $r \sim 10 \, M_0$ fourth-order convergence is found even at lower resolutions.

\begin{figure}
    \centering
    \includegraphics[width=\columnwidth]{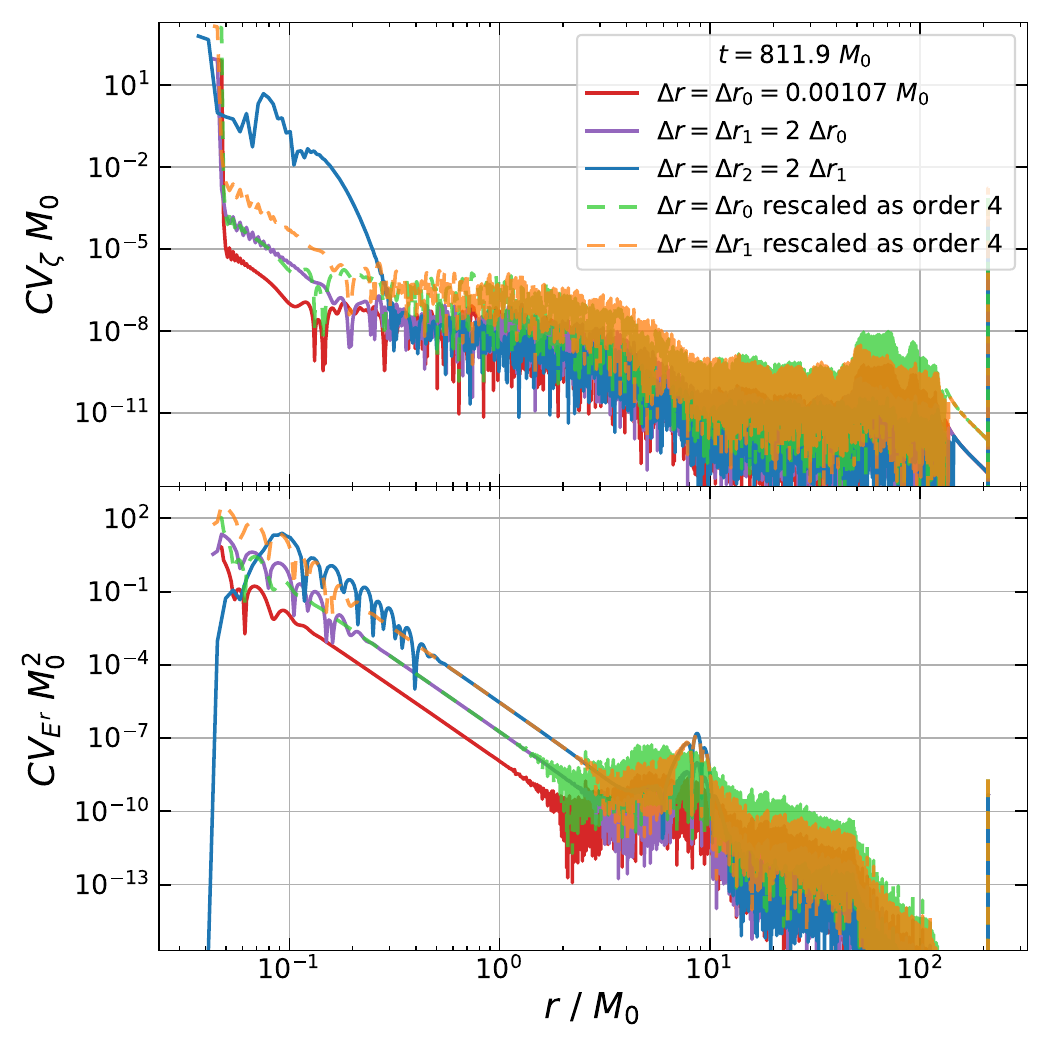}
    \caption{Scaling of the constraint violations for the simulation in which the dynamics is in the nonlinear regime, and $\alpha = 0.6$. Conventions are the same as in Fig.~\ref{fig:Alpha0.5_linear_convergence}. In this case the simulation at lowest resolution ($\Delta r = \Delta r_2$) crashed, and we evaluate the convergence at the last time step for which we have data, $t = 812 \, M_0$. Interestingly the code satisfies the expected scaling, with deviations appearing only on $CV_\zeta$ extracted from the simulation at lowest resolution.}
    \label{fig:Alpha0.6_nonlinear_convergence}
\end{figure}

Let us now focus on the simulation in the nonlinear regime. Here the run with $\Delta r = \Delta r_2$ crashed, and therefore we decided to evaluate the convergence at $t = 812 \, M_0$, which is the last time for which we have data from all the simulations. Note that the crash occurred sufficiently late, well after the appearance of the echo patterns in the behavior of $\tder \phi$ at $R = 100 \, M_0$ (cf. Fig.~\ref{fig:SimulationNonlinearAlpha0.6}). The scaling of the constraint violations is shown in Fig.~\ref{fig:Alpha0.6_nonlinear_convergence}, using the same conventions as in Figs.~\ref{fig:Alpha0.5_linear_convergence} and~\ref{fig:Alpha0.6_linear_convergence}. We can see that the code generally converges well, with significant deviations from the expected scaling appearing only in $CV_\zeta$ extracted from the simulation at $\Delta r = \Delta r_2$.

\bibliographystyle{apsrev4-1}
\bibliography{ref.bib}

\end{document}